\documentclass[%
reprint,
superscriptaddress,
 amsmath,amssymb,
 aps,
 prl,
]{revtex4-1}
\usepackage[dvipsnames]{xcolor}
\definecolor{mblue}{RGB}{0, 114, 188}
\usepackage{hyperref}
\hypersetup{backref,pdfpagemode=FullScreen,colorlinks=true,breaklinks,urlcolor=mblue,linkcolor=mblue,citecolor=mblue}
\usepackage{graphicx}
\usepackage{dcolumn}
\usepackage{bm}
\usepackage{physics}
\usepackage{braket}
\usepackage{amssymb}

\newcommand{\beginsupplement}{%
        \setcounter{table}{0}
        \renewcommand{\thetable}{S\arabic{table}}%
        \setcounter{figure}{0}
        \renewcommand{\thefigure}{S\arabic{figure}}%
        \setcounter{equation}{0}
        \renewcommand{\theequation}{S\arabic{equation}}%
     }  
\begin{document}



\title{Faster State Preparation across Quantum Phase Transition Assisted by Reinforcement Learning}





\author{Shuai-Feng Guo}
\thanks{These authors contributed equally to this work.}
\affiliation{State Key Laboratory of Low Dimensional Quantum Physics, Department of Physics, Tsinghua University, Beijing 100084, China}

\author{Feng Chen}
\thanks{These authors contributed equally to this work.}
\affiliation{State Key Laboratory of Low Dimensional Quantum Physics, Department of Physics, Tsinghua University, Beijing 100084, China}

\author{Qi Liu}
\affiliation{State Key Laboratory of Low Dimensional Quantum Physics, Department of Physics, Tsinghua University, Beijing 100084, China}

\author{Ming Xue}
\affiliation{State Key Laboratory of Low Dimensional Quantum Physics, Department of Physics, Tsinghua University, Beijing 100084, China}
\author{Jun-Jie Chen}
\affiliation{State Key Laboratory of Low Dimensional Quantum Physics, Department of Physics, Tsinghua University, Beijing 100084, China}
\author{Jia-Hao~Cao}
\affiliation{State Key Laboratory of Low Dimensional Quantum Physics, Department of Physics, Tsinghua University, Beijing 100084, China}
\author{Tian-Wei Mao}
\affiliation{State Key Laboratory of Low Dimensional Quantum Physics, Department of Physics, Tsinghua University, Beijing 100084, China}



\author{Meng Khoon Tey}
\email{mengkhoon\_tey@tsinghua.edu.cn}
\affiliation{State Key Laboratory of Low Dimensional Quantum Physics, Department of Physics, Tsinghua University, Beijing 100084, China}
\affiliation{Frontier Science Center for Quantum Information, Beijing, China}

\author{Li You}
\email{lyou@tsinghua.edu.cn}
\affiliation{State Key Laboratory of Low Dimensional Quantum Physics, Department of Physics, Tsinghua University, Beijing 100084, China}
\affiliation{Frontier Science Center for Quantum Information, Beijing, China}


\date{\today}
\begin{abstract}

An energy gap develops near quantum critical point of quantum phase transition in a finite many-body (MB) system, facilitating
the ground state transformation by adiabatic parameter change.
In real application scenarios, however, the efficacy for such a protocol is compromised by the need to balance finite system lifetime with adiabaticity, as exemplified in a recent experiment that prepares three-mode balanced Dicke state near deterministically [Y.-Q. Zou et al., Proc. Natl. Acad. Sci. U.S.A. {\bf 115}, 6381 (2018)].
Instead of tracking the instantaneous ground state as unanimously required for most adiabatic crossing, this work reports a faster sweeping policy taking advantage of excited level dynamics.
It is obtained based on deep reinforcement learning (DRL) from a multistep training scheme we develop.
In the absence of loss, a fidelity $\ge 99\%$ between prepared and the target Dicke state is achieved over a small fraction of the adiabatically required time.
When loss is included, training is carried out according to an operational benchmark, the interferometric sensitivity of the prepared state instead of fidelity,
leading to better sensitivity in about half of the previously reported time.
Implemented in a Bose-Einstein condensate of $\sim 10^4$ $^{87}$Rb atoms, the balanced three-mode Dicke state exhibiting an improved number squeezing of $13.02\pm0.20$ dB is observed within 766 ms, highlighting the potential of DRL for quantum dynamics control and quantum state preparation in interacting MB systems.
\end{abstract}


\maketitle



One celebrated hallmark for a quantum system lies at its discrete level (eigenenergy)
and associated (orthogonal) eigenstate.
According to the quantum adiabatic theorem, under slow and continuous change of a parameter,
the state of a (Hamiltonian) system stays at the level it starts with, e.g., remaining in the ground state.
The rate of parameter change has to be much less than the corresponding level spacing
in order to avoid excitations. However, gap size or level spacing at quantum critical point (QCP)
becomes diminishingly small for increasingly larger system approaching thermodynamic limit.
Constrained by finite lifetime, sweeping cannot proceed as slowly as one wishes
 in an actual experiment.
Despite the various shortcuts to adiabaticity (STA) \cite{sta2019}
for quantum state preparation \cite{takahashi2013,campbell2015,opatrny2016} demonstrated in diverse systems
 ranging from thermal \cite{du2016} to Bose-Einstein condensate (BEC) gases \cite{bason2011}
 and trapped ions \cite{richerme2013,hu2018},
they are often restricted to a few levels and augmented by 
counteradiabatic driving terms in
the Hamiltonian.
No generally applicable strategy is known for crossing
quantum phase transition (QPT) to arrive at a transformed ground state,
except for adiabatically sweeping over the finite sized gap.

\begin{figure*} [htp!]
\includegraphics[width=0.995\textwidth]{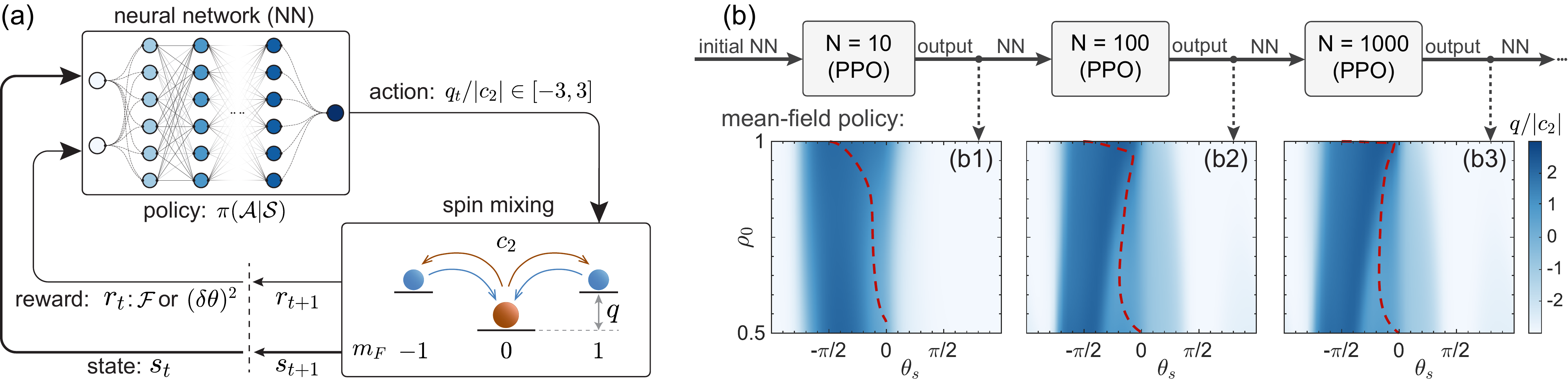}
\caption{Dicke state preparation assisted by DRL.
 (a) The learning loop between the DRL agent and the simulated system.
(b) Illustration of our multistep training scheme for successively increased system size from $N = 10$ to $N = 1000$.
Dependence of the mean values for the optimized policy $\pi^*({\cal A}|{\cal S})$ on
mean-field state space $\{\rho_0, 0, \rho_0(1-\rho_0)/2, \theta_s\}$
(mean-field policy) are displayed as density plots
in (b1), (b2), and (b3), respectively, for the corresponding $N$.
Their close resemblance reflects nice generalization ability of the policy.
Red dashed lines denote evolution trajectories starting from the polar state.
}
\label{fig1}
\end{figure*}

This Letter reports a sweeping protocol for faster crossing of QCP
benefited from excited state dynamics 
by deep reinforcement learning (DRL) \cite{bukov2018, mehta2019}. 
Beside beating the adiabatic crossing through a more sophisticated parameter tuning profile,
it is advantageous as the form of the Hamiltonian is kept without requiring counteradiabatic driving terms.
By directly learning from simulated spin dynamics,
perspectives and insights are learned for controlling many coupled spins,
significantly extending the long list of previous DRL applications in 
small systems \cite{lin2018, mehta2019,dalgaard2020,haug2019,bukov2018prb,fchen2019,bukov2018,niu2019,wang2020} 
to an interacting many-body (MB) spin system.
More specifically, from the viewpoint on the Dicke state preparation,
our work accomplishes a MB optimization (of its dynamics) with DRL,
which constitutes a key research direction in the noisy intermediate-scale quantum technology
of the near future \cite{preskill2018, doria2011, sorensen2018, xu2019,  wu2020, haine2020}.

The system we study is a $^{87}$Rb atomic BEC in the ground hyperfine $F=1$
manifold with the dominant interaction between atoms being symmetric among
spin components ($m_F=0,\pm1$). This facilitates an approximate treatment with a common spatial mode \cite{law1998,ho2000,kawaguchi2012}
and gives rise to the model Hamiltonian ($\hbar=1$ hereafter),
\begin{eqnarray}
\label{eq_ham}
H&=&\frac{c_2}{2N}{\bf L}^2-q(t) N_0,
\end{eqnarray}
with the first term describing spin exchange interaction while the rest proportional to an effective quadratic Zeeman shift (QZS) $q$,
tunable by the magnetic field and/or dressing microwave field \cite{zhao2014, luo2017}.
$N =\sum_{m_F} N_{m_F}$ ($N_{m_F} = a_{m_F}^\dag a_{m_F}$) denotes the total atom number (of all $m_F$ components)
with $a_{m_F}$ ($a_{m_F}^\dag$) the atomic annihilation (creation) operator,
and ${\bf L}=\sum_{\mu, \nu} a_{\mu}^\dag {\bf F}_{\mu\nu} a_{\nu}$ refers to the collective spin
with ${\bf F}_{\mu\nu} $ the spin-1 matrix element.
$L_z=N_{+1}-N_{-1}$ measures the system magnetization, which is conserved as is $N$ in the absence of loss.
The spin exchange interaction creates (annihilates) paired atoms in $m_F=\pm 1$
at the expense (gain) of $m_F=0$ atoms. In a single spatial mode condensate, it represents
an all-to-all interaction because
${\bf L}^2 = 2(a_1^{\dag}a_{-1}^{\dag}a_0a_0+\mathrm{h.c.})+(2N_0-1)(N-N_0) + 2N + L_z^2$.
Its strength is $c_2<0$ or ferromagnetic \cite{kempen2002,chang2005,widera2006} for $^{87}$Rb atoms.
Hence, in the absence of the QZS or $q=0$, the MB ground state
takes the maximum ${\bf L}^2=N(N+1)$ or largest $L=N$,
and is $(2N + 1)$-fold degenerate spanning the Dicke state subspace
$\{|L,L_z\rangle: L_z=-N,-N+1,\cdots ,N\}$ \cite{law1998,ho2000,kawaguchi2012}.

The balanced or the $L_z=0$ 
Dicke state
is one of the most entangled states and enables measurement precision approaching Heisenberg Limit \cite{holland1993}.
Zhang and Duan \cite{duan2013} studied its preparation 
across QPT by adiabatically sweeping from $q(t)>0$ to $q(t) =0$ at a constant rate
within a sweeping time $\tau$.
At $q/|c_2| \gg 2$, the BEC ground state is polar with all atoms in $m_F=0$,
and is denoted by $|0,N,0\rangle$ in the Fock state notation $\ket{N_{-1},N_0,N_{1}}$.
Upon adiabatically sweeping to $q=0$, the ground state smoothly evolves into the balanced Dicke state
$\ket{\psi_{\rm{Dicke}}^{(0)}}\equiv|N,L_z=0\rangle$,
after crossing QPT at $q_c/|c_2| \simeq 2$ into broken-axisymmetry (BA) phase.
The level spacing
near the QCP scales
as $\propto N^{-1/3}$ \cite{leyvraz2005, dusuel2004, duan2013, xue2018},
which erects a speed limit for crossing the QCP.

In the following, we shall first discuss how DRL agent is applied to a moderate sized
system of up to $N\simeq 10^3$ atoms in the absence of loss.
An optimized policy is found capable of preparing the balanced Dicke state 
with a theoretical fidelity ($\gtrsim 0.99$) 
using a much shorter $\tau$ than the adiabatic sweep required for by following
full quantum dynamics of Hamiltonian (\ref{eq_ham}).
Next, by modeling loss as a single atom effect, the DRL agent is retrained
with an experimental sized system ($N \sim 10^4$)
according to open system dynamics, starting with the high-fidelity policy
from without loss as prior knowledge.
The resulting profile is subsequently affirmed experimentally, leading to the improved performance we report here.
\begin{figure*}
\includegraphics[width=0.995\textwidth]{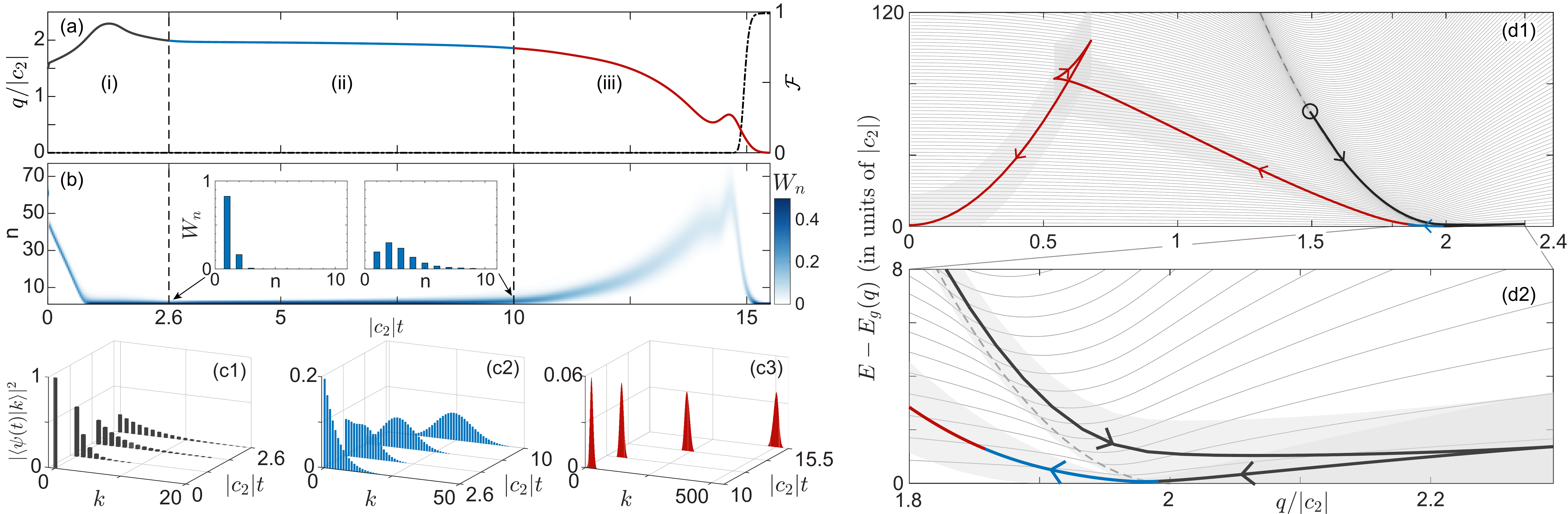}
\caption{Results from DRL policy without atom loss for $N = 2000$.
(a) The sweeping profile (in three line segments of black, blue, and red)
and corresponding simulated fidelity (dot-dashed line referred to right axis);
(b) Density plot of probability distribution $W_n = \abs{\braket{\psi_n(t)|\psi(t)}}^2$
for simulated state in the instantaneous eigenstate basis $|\psi_n(t)\rangle$ (vertical axis);
Insets show enlargements of distributions on the boundaries $|c_2|t = 2.6$ and $10$ between stages;
(c1), (c2), and (c3), respectively, show evolution of state expanded in the Fock state
basis $\ket{k,N-2k,k}$
over the three stages;
(d1) their corresponding mean energies $\expval{H(q)}{\psi(t)} - E_g(q)$ shown by the same colored
lines as in (a) with 1 standard deviation in shaded regions, and (d2) for an expanded view
around the QCP.
Arrows indicate direction of increasing time and gray dashed lines denote the CGC.
}
\label{fig2}
\end{figure*}

\textit{Reinforcement learning.---}The key task for the DRL agent is to find a time-dependent $q(t)$ profile
that prepares a state $|\psi(\tau)\rangle$ as close as possible to a target state $|\psi_{\rm Dicke}^{(0)}\rangle$
within a given $\tau$, starting from the polar state.
The learning process is as follows: In each training epoch, the sweeping $\tau$ is uniformly discretized into temporal subintervals. At each instant $t$, the agent observes the system through observable $s_t \in {\cal S}$ (by directly evaluating the expectation valuables using the current wave function) and acts by choosing $q_t \in {\cal A}$ according to the policy function $\pi ({\cal A| S})$, which gives the probability of choosing $q_i$ on observing $s_i$. The system is then evolved (according to Schr\"odinger equation) into the next instant $t+1$ using Hamiltonian (\ref{eq_ham}) with the chosen $q_t$, returns $s_{t+1}$ and a reward $r_{t}$ to the agent as illustrated in Fig. \ref{fig1}(a). This sampling process continues to the last interval, after which the collected observables and rewards in this epoch are fed into the proximal policy optimization (PPO) algorithm \cite{ppo2017} to generate a new $\pi ({\cal A|S})$ for the next epoch. The training ends when the policy converges, giving $\pi^*({\cal A|S})$ (more details can be found in the Supplemental Material \cite{supp}).

For our model system, we choose the following four observables: $\rho_0 = \braket{{N}_0}/N$, $\braket{\delta{N}_0^2}/N^2$, $|{\braket{ {a}_{+1}^{\dagger} {a}_{-1}^{\dagger}{a}_{0}^2}}|/N^2$, and $\theta_s = \arg\braket{ {a}_{+1}^{\dagger} {a}_{-1}^{\dagger} {a}_{0}^2}$ as $s_t$ \cite{state_space}.
Two types of rewards are employed, based either on fidelity between the current
and 
 the target state $\mathcal{F}=|{\braket{\psi(t)|\psi_{\rm{Dicke}}^{(0)}}}|^2$
or entanglement enhanced three-mode SU(2) interferometric sensitivity
of the current state \cite{zou2018}.
The dimension of concerned Hilbert space is $\propto N$. If DRL is directly implemented,
the large distance between the initial and the target state
points to failed training due to the sparse reward for $N\gtrsim 10^2$.
Hence, a multistep training approach is developed,
with policies from consecutively completed tasks fed forward successively to larger sized systems as illustrated in Fig. \ref{fig1}(b).
In detail, a small-sized system of $N = 10$ first trains network until $\mathcal{F} \gtrsim 0.99$,
the trained network is subsequently used as pretrained to initialize the next system
one size up, e.g., $N = 100$, leading successively to a converged policy
in the mean-field state space for larger system
size $N$ [see Figs. \ref{fig1}(b1)-\ref{fig1}(b3)]. 

The multistep training ends at $N=2000$ as subsequent training with larger system size
is beyond the computational resource we have \cite{cpu_info}.
The corresponding optimal policy achieves target Dicke state with $\mathcal{F} \simeq 0.99$
over $|c_2|\tau = 15.5$,
which is significantly shorter than the linear adiabatic sweep that requires $|c_2|\tau \gtrsim 600$,
 or a nonlinear sweep optimized for local adiabaticity that demands $|c_2|\tau \gtrsim 350$ for the same fidelity level \cite{richerme2013,laa2,supp}.

The complete evolution is found to be composed of three main stages,
partitioned by the vertical dashed lines as illustrated in Fig. \ref{fig2}.
Its understanding heavily relies on the $q$-dependent features
in the excitation spectra of Hamiltonian (\ref{eq_ham}),
many of which are shared by the broad class of Lipkin-Meshkov-Glick (LMG) model.
They exhibit a characteristic critical gap curve (CGC) \cite{supp,solinas2008},
which connects successive level spacing minima
(also scaling as $N^{-1/3}$ \cite{leyvraz2005,dusuel2004,xue2018})
and is approximately described by,
\begin{eqnarray}
\label{CGC}
E_c(q) - E_g(q) = {N(q - 2|c_2|)^2}/({8|c_2|}),~~(q/|c_2|\leqslant2),\ \ \
\end{eqnarray}
for our model as shown by gray dashed lines in Figs. \ref{fig2}(d) 
with $E_g(q)$ the ground state energy of $H(q)$. 
Sweeping quickly into either direction, the most likely trajectories for states near the CGC
will travel diabatically along the CGC in the same direction. For instance,
starting from the ground state at a high $q$ and rapidly sweeping down, the state will simply ascend along the CGC by diabatically crossing successive level spacing minima to reach increasingly higher excited states.
In other words, the initial polar ground state at high $q$ can be regarded as a superposition state in the CGC region at any $q$ (produced by an abrupt change of $q$).
As shown in Fig. \ref{fig2}(a), the starting $q$ of the DRL profile thus
sits on the left-hand-side of the QCP, or the same side as in the end after crossing QPT, to access faster spin mixing dynamics,
as otherwise starting from the right-hand-side ($q>q_c$)
would be ineffective due to insufficient excited state components. 
The excitations facilitate faster dynamics as illustrated in the following three stages:

(i) For $|c_2|t\in [0,2.6]$, this stage is denoted by the black solid line segment
starting from $q < q_c$ and marked by a circle in Fig. \ref{fig2}(d1),
during which $q(t)$ first ascends quickly 
to cross $q_c$ and arrives at $q_{\rm max}/|c_2|\simeq 2.3$ on the right-hand-side of the QCP.
The current state nearly
tracks the CGC to successively lower excited states,
but does not fall all the way down to the ground state at $q_{\rm max}$.
It is transformed
into a superposition of several low lying states due to the $q$-controlled spin-mixing  \cite{supp}.
The multilevel amplitudes during subsequent descending of $q$ to near $q_c$ can be easily controlled to
interfere constructively into the ground state level and destructively to higher levels \cite{xu2019},
which suppresses increased spreading of excitations along the CGC.


(ii) The middle interval $|c_2|t\in [2.6\emph{},10]$ (blue solid line segment) covers
the actual crossing of the QCP, where the state (from end of the first stage)
dominated by the ground and the first excited states
evolve slowly into a slightly excited form dominated by the first few low-lying levels
in the instantaneous eigenbasis $\ket{\psi_n(t)}$ [insets in Fig. \ref{fig2}(b)].
At the end of this stage $q/|c_2| \simeq 1.86$,
the state becomes a Gaussian-like wave packet
in the Fock state basis $\ket{k, N-2k, k}$ [Fig. \ref{fig2}(c2)].
It closely matches the ground state of a slightly larger $q/|c_2|\sim 1.92$,
representing a displaced Gaussian packet in a harmonic trap anticipating
for rapid translation over the next stage.
Strict adiabatic condition is observed for this stage,
reflecting the adiabatic speed limit of 
level spacing ($\sim N^{-1/3}$) near the QCP, 
in agreement with the paradigm of adiabatic Landau-Zener crossing \cite{landau1932,*zener1932,supp}.
%

(iii) During the final stage of $|c_2|t\in[10,15.5]$, the red solid line segment,
the profile corresponds to a rapid translation of the Gaussian wave packet
from the small $k$ region to balanced Dicke state at $k \sim N/4$ as shown in Fig. \ref{fig2}(c3),
accompanied by rapidly increasing fidelity shown with the black dot-dashed line in Fig. \ref{fig2}(a).
Such an overall center of mass translation
is well described by a tunable harmonic oscillator model \cite{supp,leyvraz2005},
\begin{equation}
\label{eq_hamK}
H/(N/2)= (1 - \lambda^2)(\mu - \mu_0)^2 /(2\lambda^2)-\lambda^2\varphi^2/2,
\end{equation}
with frequency $\sqrt{1- \lambda^2}$ and effective mass $1/\lambda^2$ for $\lambda = q/(2|c_2|)$,
where $\mu$ and $\varphi$ are the canonical ``position'' and ``momentum,'' respectively, and $|\mu_0| = \sqrt{1 - \lambda^2}$
denotes the center of the ground state Gaussian wave packet.
Hence, during this stage the DRL profile simply shifts the Gaussian
wave packet right after crossing the QCP at $\mu_0 \approx 0$ to $|\mu_0| \simeq 1$,
following a STA with a tunable harmonic trap \cite{sta1,sta2,sta3,sta4}.
It is accomplished by first accelerating the trap, followed by deceleration to a standstill, demonstrated by the rise and fall of the average energy [Fig. \ref{fig2}(d1)] \cite{supp}.

\textit{Experiment.---}
However, the minimal level spacing near the QCP reduces to a few Hertz for a typical $^{87}$Rb BEC with $10^4$ atoms. QPT from adiabatic level crossing therefore demands a sweeping duration $\tau$ of seconds, during which atom loss, e.g. from three-body collisions, can no longer be ignored.
In earlier preparation of balanced Dicke state ensembles \cite{zou2018}, about $5\%$ of the total atoms were lost over $\tau = 1.5$ s.
To take atom loss into account, we retrain the DRL agent by simulating the system evolution using coupled stochastic differential equations (SDEs) derived from the quasiprobability distribution based on truncated Wigner approximation \cite{steel1998, sinatra2002, norrie2006, opanchuk2012, drummond2017, johnson2017, gerving2012, hamley2012},
with atom loss modeled by one-body decay~\cite{supp}.
Instead of fidelity $\mathcal{F}$, the SU(2) interferometric sensitivity
$(\delta\theta)^2 = (3\braket{(\delta{ L}_z)^2_{\theta = 0}} + 1/2)/\braket{{\bf L}^2}$ \cite{zou2018}
for a small rotation angle $\theta$ (from the equatorial plane in the generalized Bloch sphere  \cite{lucke2011})
is used as a more suitable reward.

\begin{figure} [htp!]
\includegraphics[width=0.995\columnwidth]{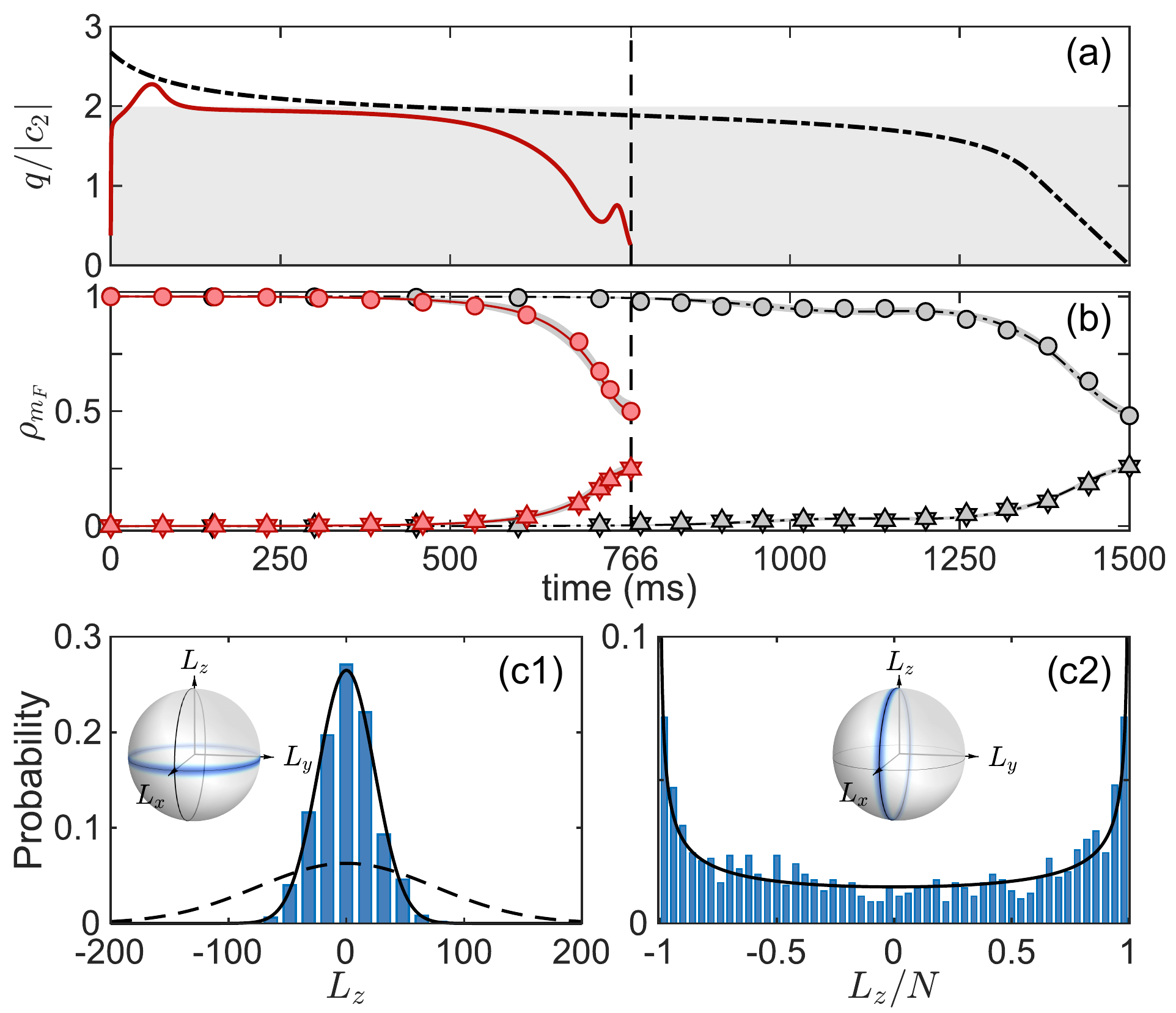}
\caption{Results (including loss) for $N \simeq 11 800$.
(a) Comparison of $q(t)$ profile from DRL (red solid line) and the
empirically optimized one from Ref. \cite{zou2018} (black dot-dashed line).
The shaded region paints the BA phase.
(b) The corresponding evolutions for fractional populations $\rho_{m_F} = N_{m_F}/N$
are compared in the same colors as in (a), with lines (gray shaded regions)
denoting simulated mean values (uncertainties).
Data points from ten experimental runs are marked by circles or triangles,
respectively for $m_F= 0$, $\pm1$ components.
The results for $m_F = \pm1$ overlay almost completely due to their correlated generation.
(c1) The measured histograms of $L_z$ for the prepared Dicke state obtained from 1000 continuous experiment runs.
The solid line denotes a Gaussian fit, whereas the dashed line is the theoretical envelope for a coherent state with same $N$.
(c2) The histograms of $L_z/N$ for the prepared state after a $\pi/2$ rotation about the $x$ axis from 1057 continuous experiment runs. The solid line represents the theoretical results for an ideal balanced Dicke state.
The inset spheres in (c) denote the generalized Bloch sphere representations for the
corresponding states.
}
\label{fig3}
\end{figure}

To keep the desirable characteristic features of the high-fidelity profile for $q(t)$, prior results trained using $\cal F$
are adopted as pre-trained instead of starting afresh DRL training using $(\delta \theta)^2$.
The network parameters from a small-sized system ($N = 100$) trained with $\cal F$
in the absence of loss are used to initialize the subsequent
network followed by solving SDEs using $(\delta\theta)^2$
as reward to successively larger systems, analogous to that illustrated in  Fig. \ref{fig1}(b).
The resulting $q(t)$ profile for $N = 11 800$ [Fig. \ref{fig3}(a)]
gives a maximal theoretical sensitivity for $(\delta \theta)^2$ at $|c_2|\tau = 13$.
It exhibits all characteristic features of the profile obtained from without loss at $N=2000$, except starting and ending at changed $q$. Our simulation shows the performance of the profile is sensitive to inaccuracies in $q/|c_2|$ due to its reliance on quantum interference, but the stringent condition for stabilizing $q/|c_2|$ is experimentally surmountable. Specifically, a variation of $q/|c_2|$ by 0.03 from the optimal profile, which is within the capability of our setup, would degrade $(\delta \theta)^2$ by only 5\%.

Experimentally implemented in a $^{87}$Rb BEC of $N \simeq 11 800$ atoms at $c_2 = - (2\pi)\times2.7~{\rm Hz}$,
the corresponding sweeping time becomes $\tau\sim766$ ms, or about half of the previously reported time \cite{zou2018} that is based on an empirical piecewise analytic profile as compared in Fig. \ref{fig3}(a). Following Ref. \cite{zou2018}, the experiment starts from preparing a BEC in the $m_F=0$ component at $q/|c_2|\sim17$, followed by ramping $q$ to 3$|c_2|$ in 300 ms before applying the DRL profile or the empirical one of Ref. \cite{zou2018}. Their corresponding evolutions of the fractional populations $\rho_{m_F}(t)$ are shown in Fig. \ref{fig3}(b).
Including detection noise and other stochastic influences,
we measure $(\delta{L}_z)_{\theta=0} = 23.88 \pm 0.52$, which gives a number squeezing of $\xi_N^2 = -20 \log_{10}[(\delta {L}_z)_{\theta=0}/\sqrt{N}] \simeq 13.02 \pm 0.20~{\rm dB}$ (vs $12.56 \pm 0.20$ {\rm dB} earlier \cite{zou2018, erratum}) below the quantum shot noise of $\sqrt{N} \simeq 106.93 \pm 0.84$ for the polar state [Fig. \ref{fig3}(c1)].
Despite reduced loss from nearly halved sweeping time,
the quality for the prepared Dicke state is only marginally improved due to
detection noise
($\delta {L}_z^{\rm DN} = 21.5$), which is dominate but quantitatively well understood \cite{zou2018}.
After taking out detection noise, we infer a number squeezing of $20.25 \pm 1.0~{\rm dB}$ (vs $17.83 \pm 0.63$ {\rm dB} earlier \cite{zou2018, erratum}).
The quality of our prepared Dicke state is further characterized by comparing directly its $L_z/N$ distribution with the ideal balanced Dicke state 
after a $\theta = \pi/2$ rotation \cite{zou2018}. The measured results are shown by the
histogram in Fig. \ref{fig3}(c2), which infers a normalized collective spin length squared of $\braket{{\bf L}^2/[N(N+1)]} = 0.998 \pm 0.021$
(see Supplemental Material \cite{supp} for all experimental details). Adopting a better detection technique, such as fluorescence detection \cite{qu2020},
would help harvesting the full advantage of the DRL policy in the future.

In summary, we develop a multistep DRL training scheme, where the agent is pretrained with small-sized systems
without loss to arrive at a high-fidelity policy, and subsequently retrained by including loss
 for larger systems to quest for enhanced interferometric sensitivity.
By adopting the consequent DRL profile to a BEC of $^{87}$Rb atoms,
the validity of our policy is affirmed based on observing improved quality Dicke state ensembles.
The optimal DRL profile includes three stages for the underline dynamics of spin mixing.
Except in the middle stage, where adiabaticity is maintained to the best possible level
constrained by the total sweeping time, faster evolutions are found in the beginning and the ending stages,
capitalizing on the features of the excited energy levels.
Our work highlights the application of DRL to guide experiments where training with
realistic system is difficult or impossible but simulation can be easily performed.
\begin{acknowledgments}
We thank Dr. L.-N. Wu, Dr. P. Xu, and Dr. Y.C. Liu for helpful discussions.
This work is supported by the Key-Area Research and Development Program of GuangDong Province
(Grant No. 2019B030330001), by the National Key R\&D Program of China (Grants
No. 2018YFA0306504 and No. 2018YFA0306503), and by the National Natural Science
Foundation of China (NSFC) (Grants No. 91636213, No. 11654001, No. 91736311, No. 91836302, and No. U1930201).
\end{acknowledgments}

\bibliographystyle{apsrev4-1}

%

\clearpage
\onecolumngrid

\begin{center}
  \textbf{\large Supplementary Material}\\[.2cm]
\end{center}
\beginsupplement

This supplementary provides expanded discussions on several issues left out of the main text due to space limitation.
They include: (i) detailed description of deep reinforcement learning (DRL) task in Sec. \ref{sec1};
(ii) generalization ability of the DRL policy obtained in Sec. \ref{sec2};
(iii) comparison among adiabatic sweeps and the DRL profile in Sec. \ref{sec3};
(iv) multilevel oscillation during the quantum critical point (QCP) crossing in Sec. \ref{sec4};
(v) comparison between full quantum simulations with mean-field simulations using the DRL profile in Sec. \ref{sec5};
(vi) connection between our model and the Lipkin-Meshkov-Glick (LMG) model in Sec. \ref{sec6};
(vii) derivation of the approximate simple harmonic oscillator model used in the third stage in Sec. \ref{sec7};
(viii) deteriorating quality of the Dicke state with atom loss in Sec. \ref{sec8};
(ix) analysis of the sensitive dependence on parameters in Sec. \ref{sec9};
and finally (x) experimental details and implementation in Sec. \ref{sec10}.

\section{the deep reinforcement learning task} \label{sec1}
The DRL task is modeled by a Markov decision process as illustrated in the main text. At each time 
instant $t$,
the agent observes $s_t \in \mathcal{S}$ of the environment and selects an action $a_t \in\mathcal{A}$
according to the current policy $\pi({\cal A|S})$. The environment then evolves to $s_{t+1}$ after the action $a_t$ and a scalar reward $r_{t}\in\mathcal{R}$
is computed and returned back to the agent. Policy $\pi$ is subsequently updated through such experience data to maximize the accumulated reward $R$.
The environment here is a simulated system with spin dynamics based on Hamiltonian (1).
The definitions for state $\mathcal{S}$, action $\mathcal{A}$, and reward function $\mathcal{R}$ in our model system of spin-1 Bose-Einstein condensate (BEC) take the following more specific meanings.

\begin{itemize}
\item $\mathcal{S}$: In the spin-1 BEC system, the set of four physical observables, including $\braket{ {N}_0}/N$, $\braket{\delta {N}_0^2}/N^2$, $|{\braket{ {a}_{+1}^{\dagger} {a}_{-1}^{\dagger} {a}_{0} {a}_{0}}}|$, and $\arg\braket{ {a}_{+1}^{\dagger} {a}_{-1}^{\dagger} {a}_{0} {a}_{0}} $ are chosen to represent the state instead of the wave function $\ket{\psi(t)}$,
    which contains complete information but is too cumbersome to handle at large $N$.
Using these more physically relevant observables as state representation facilitates directly generalizable policy due to their
explicit independence of $N$ after normalization (illustrated in the Sec. \ref{sec2}), the final policy is also easily interpreted with clear physical insights.
For our choice, $\{ \braket{ {N}_0}/N, \arg\braket{ {a}_{+1}^{\dagger} {a}_{-1}^{\dagger} {a}_{0} {a}_{0}}\}$ corresponds
 to the mean-field parameters $\{\rho_0=\braket{ {N}_0}/N, \theta_s\}$, while $\{ \braket{\delta  {N}_0^2}/N^2, |\braket{ {a}_{+1}^{\dagger} {a}_{-1}^{\dagger} {a}_{0} {a}_{0}}| \}$ calibrate first-order quantum correlations.

\item $\mathcal{A}$: The action space is a continuous range of quadratic Zeeman shift $q/|c_2| \in [-q_{\rm max},q_{\rm max}]$ with $q_{\rm max}=3$.

\item $\mathcal{R}$: Two types of object function $f(\psi)$ are employed as rewards, overlap (or fidelity) between the current and the target state $ f(\psi) = \mathcal{F} = |\braket{\psi(t)|\psi_{\rm Dicke}^{(0)}}|^2$ from full quantum simulation in the absence of loss, and $f(\psi) = \braket{(\delta \theta)^2}_{\rm min}/\braket{(\delta\theta(t_j))^2}$
    when loss is included with $(\delta\theta)^2 = [ 3\braket{\delta(L_z)^2_{\theta = 0}} + 1/2 ]/\braket{L^2_{\rm eff}}$ the expected interferometric sensitivity of current state, while $\braket{(\delta\theta)^2}_{\rm min} = 1/(2\braket{L_{\rm eff}^2})$ is the optimal (minimum value for an ideal target Dicke state $\ket{\psi_{\rm Dicke}^{(0)}}$).
The total reward is calculated as a cumulative sum of all instantaneous rewards, $R_{\rm tot} = \sum_{j=1}^{m} r_j$ with $r_j = f(\psi(t_j)) - f(\psi(t_{j - 1}))$.
Starting with the initial (polar) state, the object function satisfies $f(\psi_i) = 0$, while $f(\psi_f) \simeq 1$ denotes a
complete success with perfect fidelity or sensitivity in the end.
Such a dense cumulative partition of reward makes the training process quicker and more stable.
In addition, the instantaneous reward can be further revised into
\begin{equation}
\label{ }
r_j \rightarrow \log_{10} \Big( \frac{1 - f(\psi(t_{j-1}))}{1 - f(\psi(t_{j}))}\Big),
\end{equation}
to prevent deteriorating learning efficiency when the state approaches the target ($f(\psi(t))\rightarrow 1$).
\end{itemize}

\begin{figure}[!htp]
\begin{center}
\includegraphics[width=0.95\textwidth]{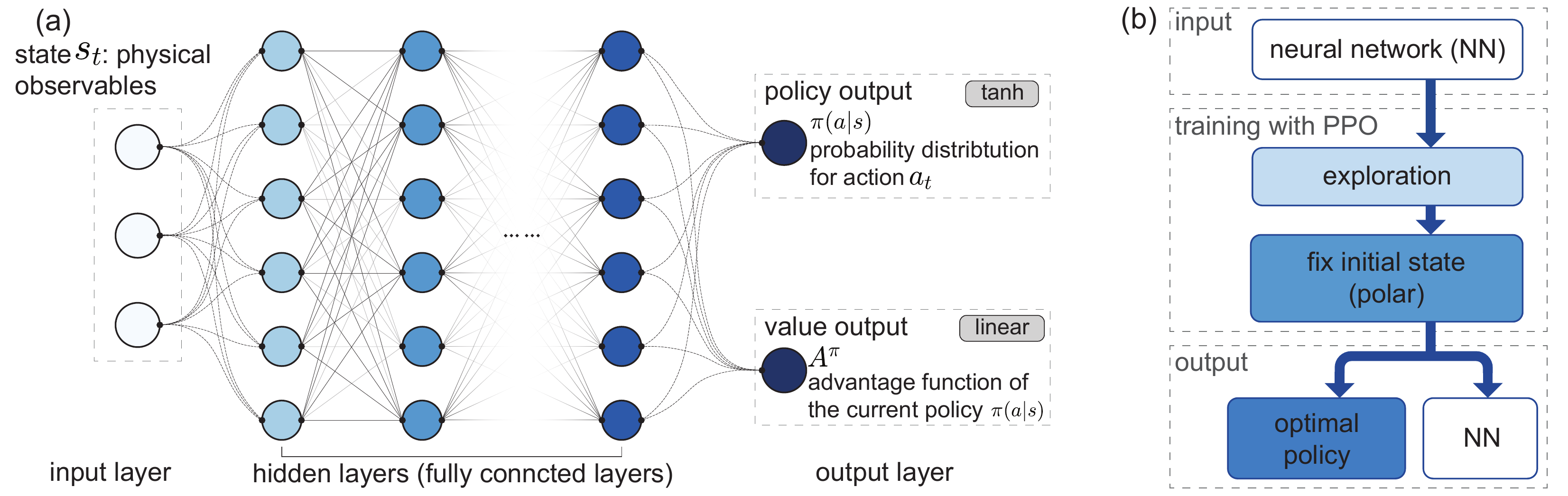}
\caption{(a) Illustration of the neural network structure for the PPO algorithm. The hidden layers contain four fully connected ones with $[64,32,16,8]$ neurons respectively. (b) Illustration of the learning process at a specific system size $N$. Either a randomly initialized network or a trained network can be successively used as input network to the next one.}
\label{fig_sup1}
\end{center}
\end{figure}

The proximal policy optimization (PPO) algorithm is employed to find the optimized policy $\pi^*$ that maximizes cumulative reward $R$,
\begin{equation}
\label{ }
\pi^* = \mathop{\arg\max}_{\pi} R, ~~\mathrm{with} ~~ R=\sum_j\gamma^tr_j,
\end{equation}
and $\gamma$ a discount factor, which is typically chosen very close to 1 to avoid greedy solutions.
The structure of the neural network (NN) for PPO algorithm is shown in Fig. \ref{fig_sup1}(a)
and the pseudo-code for PPO algorithm is shown in Table \ref{PPO_pseudo_code}.
The PPO algorithm we employed for this work comes from the OpenAI SpinningUp library (TensorFlow version) \cite{ppo}.
To facilitate the training, we encapsulate the quantum state evolution into a Gym environment as suggested by OpenAI \cite{gym}.
With PPO algorithm, the policy is stochastic which returns a normalized distribution on action space for a given state $s$ and satisfies
\begin{eqnarray}
\int_{a \in \mathcal{A} }\pi(a|s) {\rm d}a \equiv 1.
\end{eqnarray}

When a policy is reached, one can either choose the action with the maximum probability $a=\max_{a'}\pi^*(a'|s)$ as a deterministic protocol
or select the best one among multiple profiles sampled based on $\pi^*(a'|s)$.
The former approach is taken by us in this work, which leads to the DRL profile $q(t)$ of the policy.

For a system of $N$ atoms, a deep NN is adopted to parameterize the actor and the critic networks in PPO,
each containing four fully connected hidden layers with $[64,32,16,8]$ neurons respectively.
Every learning episode is divided into a few hundred consecutive steps and the total evolution time is limited to $\tau$
that depends on the total number of atoms $N$. Other hyper-parameters used in the training are listed in Table \ref{table-ss-1},
some of them are tuned within a range according to system size.
Typically, thousands of training epochs are required to reach an optimal policy,
while each training epoch contains hundreds of learning episodes.
To accelerate the training process and enhance the resulting performance of the optimal policy, the total training epochs are divided into two parts as illustrated in Fig. \ref{fig_sup1}(b).
For the first few hundred epochs, a random quantum state is used as an initial state to ignite a learning episode, which helps the agent to learn the basic geography of the state space (exploration for short). Afterwards, the initial state is reset to polar state $\ket{0,N,0}$
and the agent subsequently finds out a (sub)optimal controlled trajectory in state space $\cal S$ from the polar state to the target (Dicke) state.

Due to the almost independence on $N$ of the physical observables in state $s_t$, training tasks in systems of different $N$ share the same NN structure, i.e., a larger-sized system inherits the trained NN from a smaller-sized system as a pre-trained network, which dramatically reduces the required number of training epoches in the larger-sized system,
and the total training process becomes efficient since training in larger-sized systems no longer consumes enormous computational resources.
Such a multi-step training process is adopted for training increased system size
from $N = 10$ to $N = 2000$ atoms using full quantum simulation of
Schr\"odinger equation without loss,
and from $N = 1000$ to $N = 11800$ atoms by 
following stochastic differential equations from the
corresponding master equation including atom loss modeled by one-body decay.

\begin{table}[!htp]
\centering
\caption{Pseudo-code of PPO algorithm \cite{ppo2017}}
\begin{tabular}{l}
\hline
PPO algorithm \\
\hline
1. Input: initial weights of policy network $\theta_0$, initial weights of value function $\phi_0$ \\
2. $\bold{for}$ k = 0, 1, 2, ... $\bold{do}$ \\
3. ~~~~~~Collect multiple trajectories of spin state evolution $\mathcal{D}_k=\{\tau_i\}$ under current policy $\pi_{\theta_k}$ \\
~~~~~~~~~~in self-defined Gym environment. \\
4. ~~~~~~Computes rewards-to-go $ {R}_t=\sum_{l=0}^{|c_2| t_c-t-1}\gamma^l r_{t+l}+V_{\phi_k}(s_{|c_2| t_c})$. \\
5. ~~~~~~Use GAE-$\lambda$ method \cite{gae2018} and current value function $V_{\phi_k}$ to estimate advantage function $A^{\pi_k}$. \\
6. ~~~~~~Maximize PPO-Clip lower bound function and update weights of policy network \\
~~~~~~~~~\qquad\qquad\qquad $\displaystyle \theta_{k+1}=\mathop{\arg\max}_{\theta_k} \dfrac{1}{|\mathcal{D}_k||c_2| t_c}\sum_{\tau\in\mathcal{D}_k}\sum_{t=0}^{|c_2| t_c} L_k^{\text{clip}}(s_t,a_t)$, \\
~~~~~~~~~~usually using gradient descents methods such as Adam \cite{adam2017} and SGD \cite{sgd1951,sgd1952}. \\
7. ~~~~~~Minimize mean-squared error and update weights of value function \\
~~~~~~~~~\qquad\qquad\qquad $\displaystyle\theta_{k+1}=\mathop{\arg\min}_{\phi_k} \frac{1}{|\mathcal{D}_k| |c_2| t_c}\sum_{\tau\in\mathcal{D}_k}\sum_{t=0}^{|c_2| t_c} \left( V_{\phi}(s_t)- {R}_t \right)^2$, \\
~~~~~~~~~~usually using gradient descents methods. \\
8. $\bold{end~for}$ \\
\hline
\end{tabular}
\label{PPO_pseudo_code}
\end{table}

\begin{table}[!htp]
\caption{Training Hyper-parameters for PPO in SpinningUp \cite{ppo}}
\centering
\begin{tabular}{rl}
\hline
Hyperparameters& \hskip 6pt value\\
\hline
hidden size& \hskip 6pt [64, 32, 16, 8] \\
activation& \hskip 6pt $\tanh$ \\
discounted factor $\gamma$& \hskip 6pt 0.99 \\
actor-network learning rate& \hskip 6pt 3E-4 \\
critic-network learning rate& \hskip 6pt 5E-4 \\
steps/episode& \hskip 6pt  $1000\sim20000$ \\
target KL-divergence& \hskip 6pt $0.01\sim0.1$ \\
clip ratio $\epsilon$& \hskip 6pt 0.2  \\
GAE-$\lambda$& \hskip 6pt 0.97  \\
\hline
\end{tabular}
\label{table-ss-1}
\end{table}


\section{Generalization ability of the DRL policy} \label{sec2}
A DRL policy is typically trained at a specific system size with $N$ atoms. When observables selected for the DRL agent are $N$-independent or almost $N$-independent as in our task, an essentially $N$-independent policy 
will be learnt.
The performance of such a policy when applied to different training conditions, e.g., different numbers of atoms $\tilde{N}$,
different sweeping time $\tilde{\tau}$, or different range of action space $[-\tilde{q}_{\rm max},-\tilde{q}_{\rm max}]$,
is measured by generalization ability of a policy.
Here, we focus on the generalization for the number of atoms $N$
 as well as the sweeping time $\tau$ since we hope to obtain a policy
that can handle larger systems and prepare the target state within a shorter time.

\begin{figure}[!htp]
\begin{center}
\includegraphics[width=0.9\textwidth]{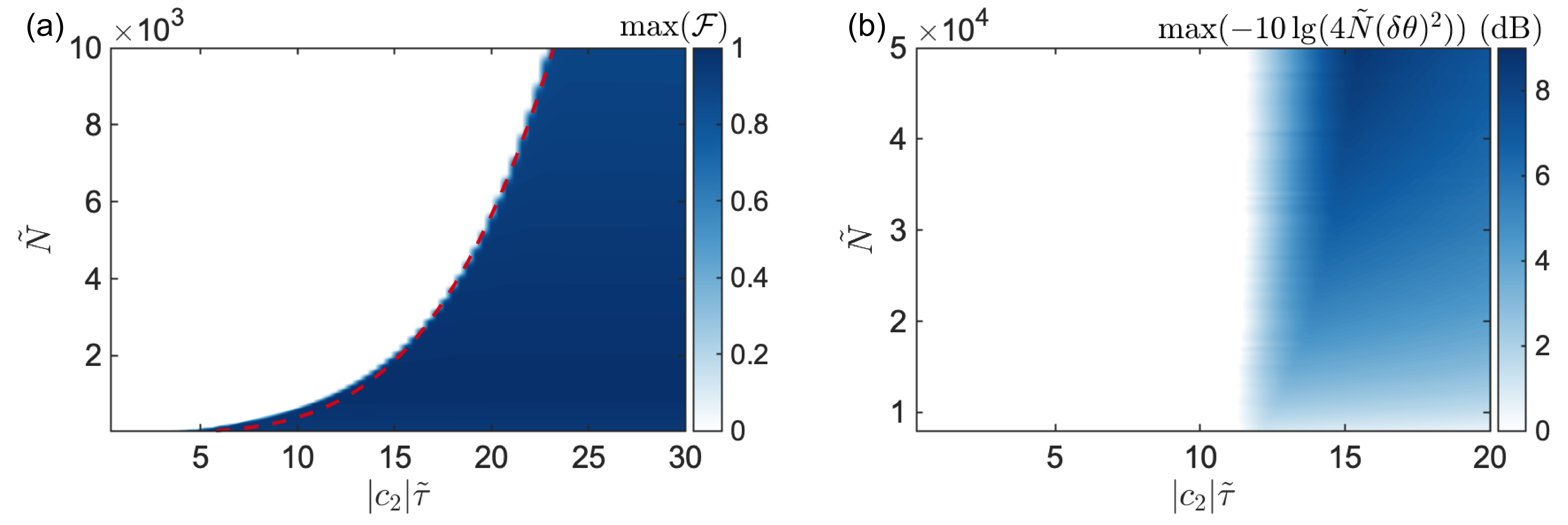}
\caption{Illustrating generalization ability for (a) the $N = 2000$ fidelity policy (without loss)
at $|c_2|\tau = 15.5$ to $\tilde{N}\in [50,10000]$ and $|c_2|\tilde{\tau} \in [0.4, 30]$, and (b) the $N = 11800$ sensitivity policy (including loss and detection noise) at $|c_2|\tau$=13 to $\tilde{N}\in [8000,50000]$ and $|c_2|\tilde{\tau} \in [0.2, 20]$. 
Here, $N$ and $\tau$ are the number of atoms and sweeping time for the training passage, while $\tilde{N}$ and $\tilde{\tau}$ are those for testing passage.
The sharp boundary (red dashed line) in (a) is fitted to $|c_2|\tilde{\tau} \propto \tilde{N}^{1/\alpha}$ with $\alpha \simeq 3.8$, approximately reflecting the speed limit constrained by level spacing ($\sim1/\tilde{N}^{1/3}$) near the QCP
in the second stage of our DRL profile as analyzed in the main text. The deteriorating quality as $\tau$ increases in (b) comes from
increased influence of atom loss, which will be discussed in Sec. \ref{sec5}.}
\label{fig_sup2}
\end{center}
\end{figure}

Figure \ref{fig_sup2} illustrates generalization ability for the $N=2000$ policy using $\mathcal{F}$ as reward with $|c_2|\tau = 15.5$ (a), while (b) refers to the case of $N=11800$ policy
from including atom loss and detection noise using $(\delta\theta)^2$ as reward with $|c_2|\tau = 13$.
Both policies remain effective and perform well for system sizes $\tilde{N}$ smaller than the trained $N$.
For increasingly larger system sizes $\tilde{N}$ far above the trained $N$, however, the performance level gradually
tails off and re-training becomes necessary in order to maintain the same calibre of performance.
The situation for total sweeping time $\tau$ is similar with the performance level gradually
deteriorating when the sweeping time $\tilde{\tau}$ becomes much shorter than the trained $\tau$.
More quantitatively, we find the performance degrades rather slowly as $\tilde{N}$ increases.
The $N = 2000$ policy can achieve a target fidelity of $\mathcal{F}\gtrsim0.85$
for a system with an order of magnitude more atoms, e.g. $\tilde{N}=10^4$,
illustrating the excellent level of generalization ability with $N$.

Due to the large effective range in $N$, the multi-step training scheme we develop as described in the main text as well as in the previous section can facilitate enlarged system sizes efficiently until the limit of computation resources is approached. Similar to enlarge system size, we can also employ multi-step training to shorten sweeping time $\tau$, according to the previous analysis of generalization ability.

\section{Comparing adiabatic sweeps with DRL profile} \label{sec3}
\begin{figure}[!htp]
\begin{center}
\includegraphics[width=0.95\textwidth]{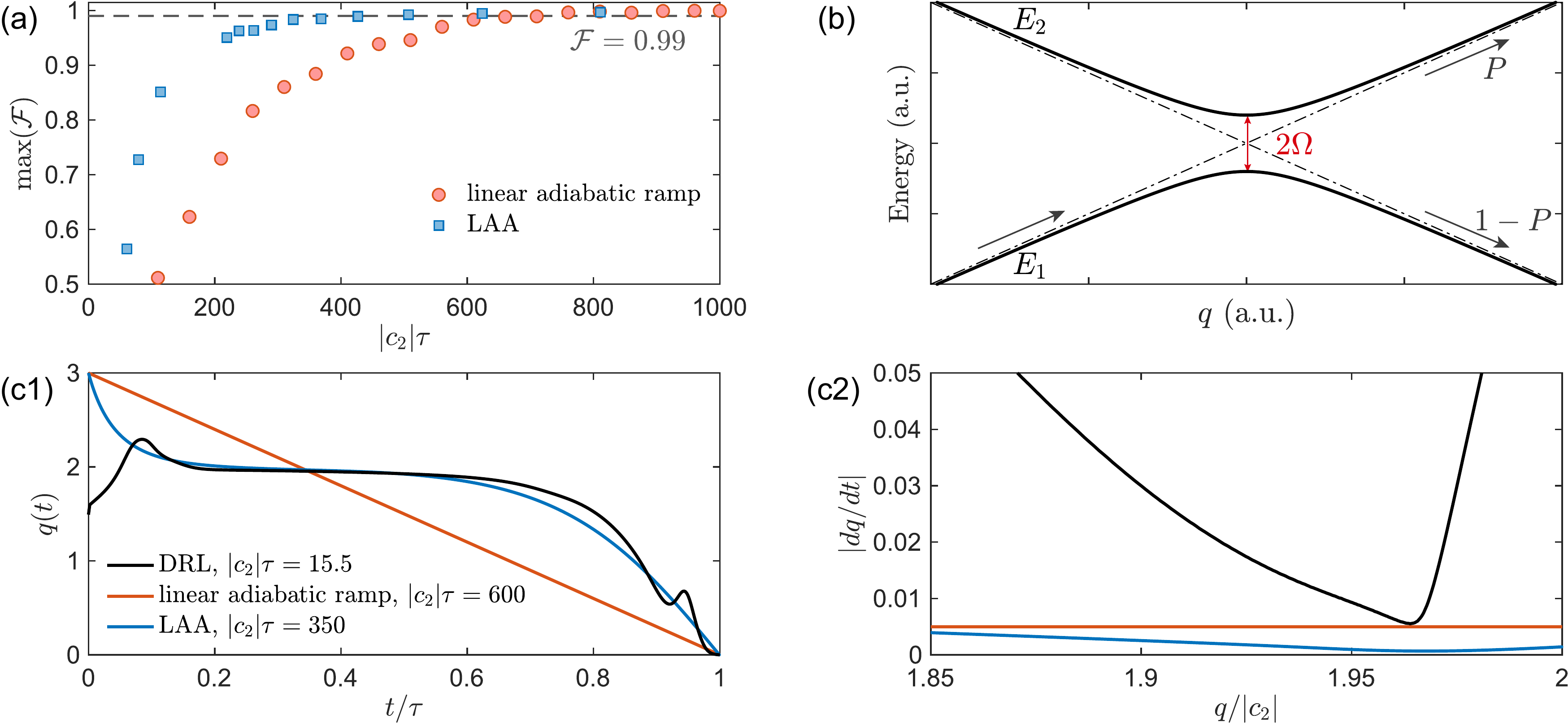}
\caption{(a) The dependence of maximum achievable fidelity on sweep time for linear adiabatic sweep (red circles) and non-linear sweep with LAA (blue squares). 
The initial state for these adiabatic sweeps is the ground state of $q/|c_2| = 3$ very close to the initial polar state (the ground state of $q/|c_2| \gg 2$) used in the DRL sweep profile.
The dashed line denotes achievable fidelity level of the $N = 2000$ DRL profile.
(b) Illustration of the Landau-Zener transition model in a two-level system of linear level crossing.
(c1) Comparing the DRL sweep profile (black dashed line) at $|c_2|\tau = 15.5$ with linear adiabatic ramp at $|c_2|\tau = 600$ (red solid line) and LAA ramp with $|c_2|\tau = 350$ (blue solid line).
Their corresponding sweep speeds around the QCP are shown in (c2).}
\label{fig_sup3}
\end{center}
\end{figure}

In this section, we compare two types of adiabatic sweep with the DRL profile for $N = 2000$ discussed as an example in the main text.
According to quantum adiabatic theorem, a quantum state of a system follows the eigen-state
it starts with under slow and continuous parameter change. For the model system we consider,
starting from the polar ground state at large $q>0$, sweeping $q$ down to $q = 0$
slowly transforms adiabatically
the ground state into balanced target Dicke state. 
At $N = 2000$, we find numerically the maximum achievable fidelity for linear adiabatic sweep
approaches the performance of DRL profile ($\mathcal{F} \gtrsim 0.99$) at a sweep time
$|c_2|\tau \gtrsim 600$ [red circles in Fig. \ref{fig_sup3}(a)].
For non-linear sweeping satisfying local adiabatic approximation (LAA) \cite{richerme2013,laa2},
the sweep time reduces to $|c_2|\tau \gtrsim 350$ [blue squares in Fig. \ref{fig_sup3}(a)].

In the paradigm two-level Landau-Zener transition model, the diabatic transition probability is given by,
\begin{equation}
\label{eq_lz}
P = \exp(-\frac{2\pi |\Omega|^2}{|\frac{dq}{dt}\frac{\partial}{\partial q}(E_2 - E_1)|}),
\end{equation}
when parameter $q(t)$ is swept across the avoided level crossing, with
$\Omega$ denoting the Rabi frequency for two state coupling, $E_1(q)$ and $E_2(q)$ the parameter
$q$-dependent eigen-energies of the two states (as illustrated in Fig. \ref{fig_sup3}(b)).
Mapped to our model, the Rabi frequency near the QCP becomes approximately $\Omega \propto N^{-1/3}$, essentially the gap size.
As $\partial(E_2 - E_1)/\partial q$ is determined by the specific $q$-dependent Hamiltonian,
the diabatic transition rate is dominantly decided by sweep speed $|\partial q/\partial t|$.
For a linear adiabatic sweep, its constant sweeping speed is determined by the minimum energy gap $\Delta_E$ between ground and the first excited state, whereas the sweeping speed for non-linear sweep satisfying LAA
is proportional to $\Delta_E^2(q)$.
The DRL sweep profile we obtain is shown in Fig. \ref{fig_sup3}(c1) and (c2). Indeed
it exhibits a sufficiently slow speed near the QCP region with which both the linear adiabatic sweep
and the non-linear LAA sweep \cite{richerme2013, laa2} can achieve
a high-fidelity target state, i.e., satisfying overall adiabaticity.
This section of adiabatic sweep in the vicinity of the QCP 
thus constitutes an unavoidable route one must overcome in order to
connect the reconfigured state after the first stage of DRL profile to the coherent state like Gaussian wave packet in the Fock basis near the end of the middle stage of DRL profile as analyzed in the main text.
The sweep speed slows down dramatically at $q \lesssim q_c$ to avoid excitation into higher energy states
in the second stage.

\section{multi-level oscillation during crossing the QCP} \label{sec4}
\begin{figure}[!htp]
\begin{center}
\includegraphics[width=\textwidth]{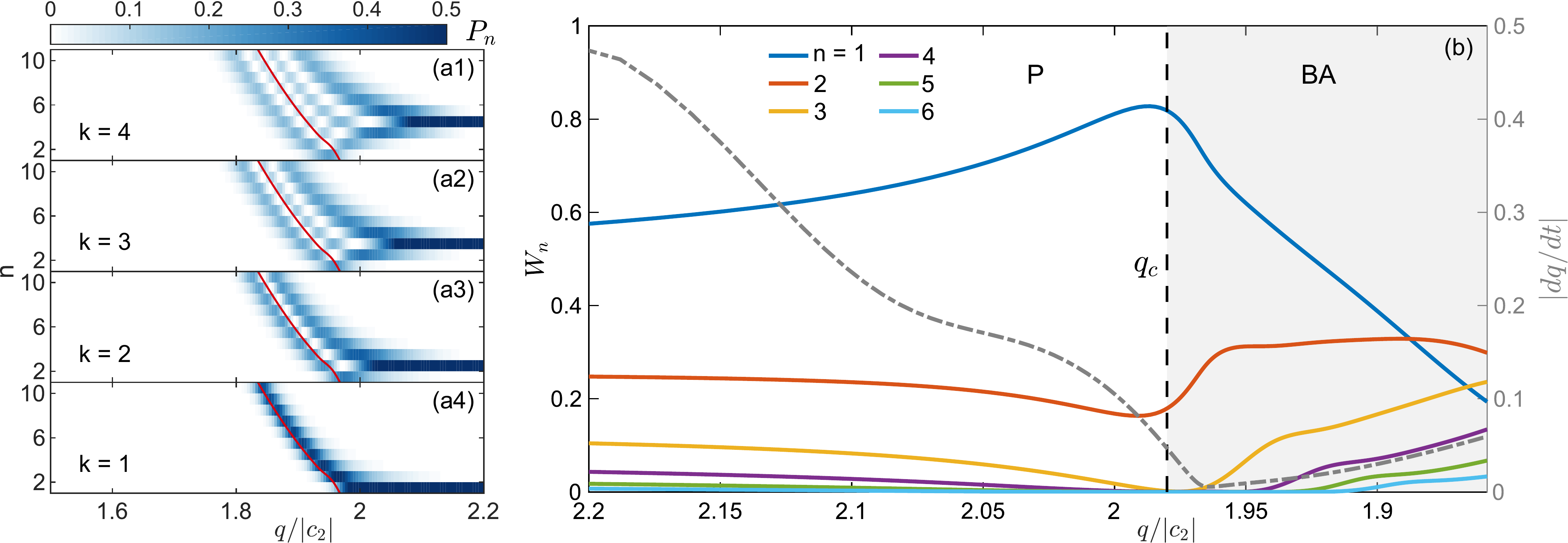}
\caption{ (a) Projections to the ground and the first three excited states of the eigen-state basis at $q/|c_2| = 2.2$
: $P_n=|\braket{\psi_n(q)|\psi_k(q/|c_2|=2.2)}|^2~(k = 1,2,3,4)$
for small $q$ at $N = 2000$. The red solid line denotes numerically calculated CGC.
(b) Evolution of the simulated probability distribution $W_n = |\braket{\psi_n(q(t))|\psi(q(t))}|^2$ following the $N = 2000$ DRL profile for the first six eigen-states in the instantaneous eigen-state basis $\ket{\psi_n(q(t))}$.
The dot-dashed line represents the sweep speed of the DRL $q(t)$ profile.
The vertical dashed line denotes the QCP which separates polar (P) phase ($q > q_c$) and broken-axisymmetry (BA) phase ($q < |q_c|$).}
\label{fig_sup4}
\end{center}
\end{figure}

Typically, with a diabatic $q$-sweep starting at an eigen-state on the right hand side of $q > q_c$, excitations are inevitable as $q$ comes down rapidly close to the QCP. The state projections onto the first four eigen-states at $q/|c_2| = 2.2$ illustrates clearly the excitation channels in the critical gap curve (CGC) region (as shown in Fig. \ref{fig_sup4}(a)).
These excitation channels can cancel out through controlled multi-level destructive interference, e.g. the channels of the first and the second excited states null out such that the probability of staying in the ground state increases.
This is akin to Rabi oscillation in a two-level system, where an initial superposition state 
can be timed to end up in either one of the two levels.
Hence, instead of the difficult and time-consuming mission of staying at the ground eigen-state
 during sweeping down from $q > q_c$, the DRL policy chooses to start at a superposition state
of low-lying levels in the first stage before crossing the QCP, which is derailed from the CGC.
Focusing on the DRL sweep starting from $q/|c_2| = 2.2$, where the state is dominated
by the first five levels, it is known in this case excited level populations can be almost suppressed completely
via multi-level oscillation with a quench-$q$ scheme \cite{xu2019}.
In our case, DRL finds a similar mechanism in action, significantly reducing excited state populations
via a sophisticated control of sweep speed $|\partial q/\partial t|$ (as shown in Fig. \ref{fig_sup4}(b)).
The state before entering the QCP region ends up being dominated by the first two energy levels.

After crossing the QCP, an analogous $q$-controlled dynamical process allows for populating the first few excited levels. Hence, a nearly displaced ground state (a displaced Gaussian wave packet in an approximately simple harmonic trap) is prepared at the end of the second stage, facilitating a rapid translation into the target state with high-fidelity.

\section{comparison between full quantum and mean-field simulations under for DRL profile} \label{sec5}

\begin{figure}[!htp]
\begin{center}
\includegraphics[width=\textwidth]{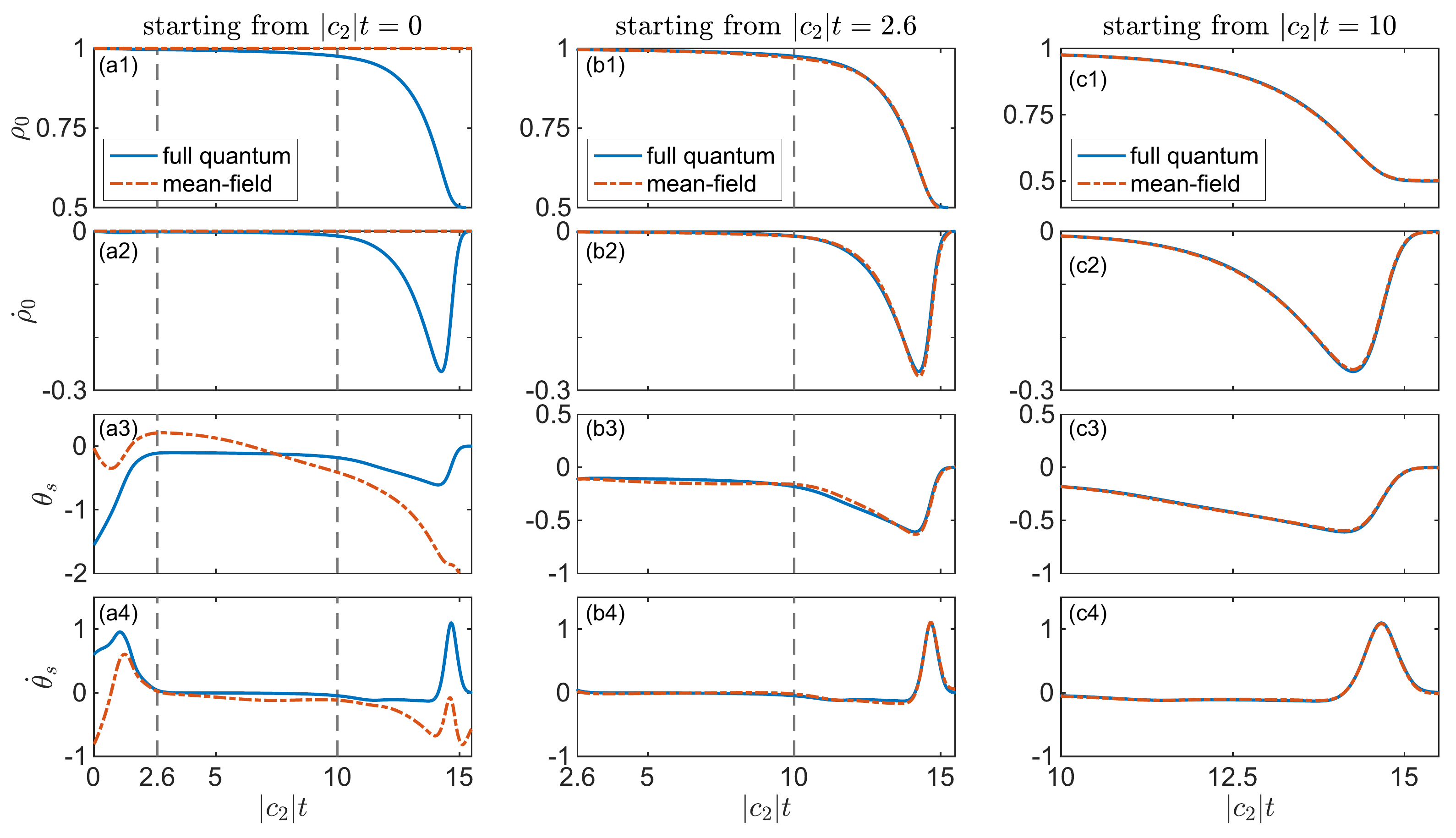}
\caption{Full quantum simulation (blue solid line) compared with mean-field results (red dot-dashed line) for the $N = 2000$ DRL profile, starting from three different instants: (a) $|c_2|t = 0$, (b) $|c_2|t = 2.6$, and (c) $|c_2|t = 10$, respectively corresponding to the
three beginning instants of the three stages in a complete DRL sweep profile.
The dashed lines mark boundaries of the three stages.}
\label{fig_sup5}
\end{center}
\end{figure}

Besides the full quantum simulation governed directly by the model Hamiltonian
which includes the quantum fluctuation,
mean-field dynamics neglecting quantum fluctuations is employed for large $N$ by
replacing annihilation (creation) operators with complex numbers:
$a_{m_F} \rightarrow \sqrt{N}\zeta_{m_F}$, $a^{\dagger}_{m_F} \rightarrow \sqrt{N}\zeta^*_{m_F}$, with $\zeta_{m_F} = \sqrt{\rho_{m_F}}e^{i\chi_{m_F}}$.
Within the $L_z = 0$ subspace, this leads to the following coupled equations
\begin{align}
\label{}
i\dot{\zeta}_1& = c'(|\zeta_1|^2\zeta_1 + |\zeta_0|^2\zeta_1  - |\zeta_{-1}|^2\zeta_1  + \zeta_0^2 \zeta_{-1}^*),   \\
i\dot{\zeta}_0& = -q(t)\zeta_0 + [(|\zeta_1|^2 + |\zeta_{-1}|^2)\zeta_0 + 2\zeta_1\zeta_{-1}\zeta_0^*],   \\
i\dot{\zeta}_{-1}& = c'(|\zeta_{-1}|^2\zeta_{-1} + |\zeta_0|^2\zeta_{-1}) - |\zeta_1|^2\zeta_{-1} + \zeta_0^2\zeta_1^*,
\end{align}
with $c' = c_2/N$. They are further simplified into two independent equations making use of the normalization condition $\sum_{m_F = -1,0,1}|\zeta_{m_F}|^2 = 1$ and the conserved magnetization $|\zeta_1|^2 = |\zeta_{-1}|^2$. Setting spinor phase $\theta_s = \chi_{1} + \chi_{-1} - 2\chi_{0}$, the coupled equations reduce to
\begin{eqnarray}
\label{eqs_meanfield}
   \dot{\rho}_0 &=& 2c'\rho_0(1-\rho_0)\sin\theta_s,   \nonumber\\
   \dot{\theta_s} &=& -2q(t) + 2c'(1 - 2\rho_0)(1 + \cos\theta_s),
\end{eqnarray}
which behaves as the non-rigid pendulum found earlier \cite{zhang2005}.

Starting in polar state with $\braket{N_{1}}/N = \braket{N_{-1}}/N = 0$, the initial evolutions
for $m_F=\pm 1$ components are dominated by quantum fluctuations,
which are neglected unfortunately in mean-field theory.
Hence, in the beginning near $|c_2|t = 0$, mean-field evolution
(red dot-dashed line in Fig. \ref{fig_sup5}(a)) is totally different
from full quantum simulation (blue solid line).
The situation changes from the beginning of the second stage as shown
in Fig. \ref{fig_sup5}(b), where evolutions begin to agree with each other
although differences remain in their detailed structures,
especially concerning $\theta_s(t)$ and $\dot{\theta}_s(t)$.
However, beginning with the third stage as shown in Fig. \ref{fig_sup5}(c),
evolutions from the above two methods match
perfectly with each other as the mean fields are well established.
Since $\psi(|c_2|t = 10)$ is already nearly an ideal
coherent state like Gaussian wave packet in the Fock basis, its dynamics is well described semi-classically
by the mean-filed approach. Hence, the control problem for the DRL agent at this stage is analogous to
the control of a classical non-rigid pendulum.

\section{connection between our model and the Lipkin-Meshkov-Glick model} \label{sec6}
The spin-1 system we consider is fully described by the SU(3) symmetry group and its correspondingly generated Lie algebra. The infinitesimal generators we employ are the Cartesian dipole-quadrupole decomposition of the Lie algebra su(3) \cite{di2010}. This leads to three dipole (or angular momentum) operators $L_a$, and nine quadrupole operators $N_{ab}$ which are moments of the quadrupole tensor $(\{a, b\} \in \{x, y, z\})$ \cite{di2010, carusotto2004, hamley2012}.

If one is concerned mainly in one of the SU(2) subspaces, $\{ L_x, N_{yz}, (N_{zz} - N_{yy})\}$, of the SU(3) group, with
\begin{align}
\label{}
    L_x & = \frac{1}{\sqrt{2}}((a_1 + a_{-1})a_0^{\dagger} + (a_1^{\dagger} + a_{-1}^{\dagger})a_0),   \\
    N_{yz} & = \frac{i}{\sqrt{2}}((a_1 + a_{-1})a_0^{\dagger} - (a_1^{\dagger} + a_{-1}^{\dagger})a_0), \\
    N_{zz} - N_{yy} & = N - 3N_0 + a_1^{\dagger}a_{-1} + a_{-1}^{\dagger}a_1,
\end{align}
it will satisfy the SU(2) commutation relationship,
\begin{equation}
\label{ }
\Big[\frac{L_x}{2},\frac{N_{yz}}{2} \Big] = i\frac{N_{zz} - N_{yy}}{4}, ~~
\Big[\frac{N_{yz}}{2}, \frac{N_{zz} - N_{yy}}{4} \Big] = i\frac{L_x}{2}, ~~
\Big[\frac{N_{zz} - N_{yy}}{4}, \frac{L_x}{2}\Big] = i\frac{N_{yz}}{2}. ~~
\end{equation}
A direct analogy to the collective spin operators $\{J_x, J_y, J_z\}$ in the LMG model can be made,
where $J_{\alpha} = \sum^{\cal N}_{k = 1}\sigma_{\alpha}^{(k)}/2$ with
 $\sigma^{(k)}_{\alpha}~(\alpha \in \{ x,y,z \})$ the corresponding Pauli matrices for the $k$-th atom.
Hence, there exists a unitary transformation $\cal U$, which maps $\{ L_x/2, N_{yz}/2, (N_{zz} - N_{yy})/4\}$ onto $\{J_x, J_y, J_z\}$, with $L_x/2 = {\cal U}^{\dagger}J_x{\cal U}$,~$N_{yz}/2={\cal U}^{\dagger}J_y{\cal U}$,~and $(N_{zz} - N_{yy})/4={\cal U}^{\dagger}J_z{\cal U}$. This unitary operator thus transforms the LMG Hamiltonian
\begin{eqnarray}
\label{eq_hamLMG}
H_{\rm LMG} = \frac{1}{\cal N} (\gamma_x J_x^2 + \gamma_y J_y^2) + h(t)J_z.
\end{eqnarray}
into
\begin{equation}
\label{eq_ham_r}
\tilde{H} = {\cal U}^{\dagger}H_{\rm LMG}{\cal U} = \frac{1}{\cal N}\Big(\gamma_x\frac{L_x^2}{4} + \gamma_y\frac{N^2_{yz}}{4}\Big) + h(t)\frac{(N_{zz} - N_{yy})}{4}.
\end{equation}

For an arbitrary state $\ket{.}$ in the zero magnetization ($L_z = 0$) subspace,
\begin{equation}
\label{ }
\braket{L_x^2} = \braket{L_y^2} = \braket{[(N_1 + N_{-1})(2N_0 -1) + 2a_1^{\dagger}a_{-1}^{\dagger}a_0a_0 + 2a_1a_{-1}a_0^{\dagger}a_0^{\dagger}]/2 + N} = \braket{{\bf L}^2}/2,
\end{equation}
and $\braket{N_{zz} - N_{yy}} = \braket{N - 3N_0} \simeq -2N_0$ with the undepleted approximation $N \simeq N_0$. Hence, if we set $\gamma_x/{\cal N} = 4c_2/N,~\gamma_y = 0,~h(t) = 2q(t)$, and focus on the $L_z = 0$ subspace, the Hamiltonian (\ref{eq_ham_r}) can be simplified into,
\begin{equation}
\label{ }
\tilde{H} = \frac{c_2}{2N}{\bf L}^2 - q(t)N_0,
\end{equation}
which is exactly the Hamiltonian of our spin-1 system.

\section{An approximate simple harmonic oscillator model in the third stage} \label{sec7}
\begin{figure}[!htp]
\begin{center}
\includegraphics[width=0.8\textwidth]{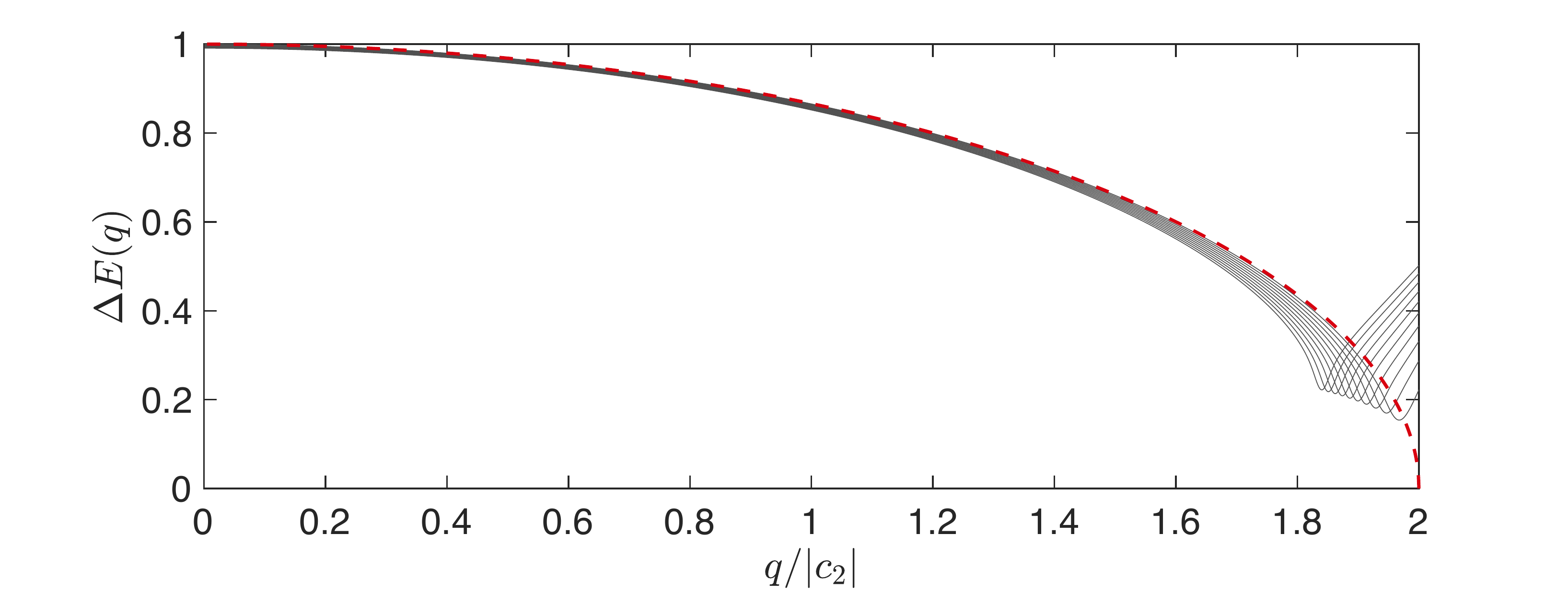}
\caption{Comparing the nearest neighbour level spacing for the $q$-dependent eigen-energies $\Delta E_j= E_{j+1}(q)-E_j(q)$ (gray lines)
for the first ten low-lying levels with the approximate harmonic frequency $\sqrt{1 - \lambda^2}$ (red dashed line) of Hamiltonian (\ref{eq_hamK2}) at $N = 2000$. The mismatch near the QCP probably comes from the finite size effect.}
\label{fig_sup6}
\end{center}
\end{figure}

As mentioned in the previous section, in the $L_z = 0$ subspace and the undepleted approximation $N \simeq N_0$,
our spin-1 Hamiltonian can be described through $\{L_x, N_{yz}, N_{zz}-N_{yy}\}$ as,
\begin{equation}
\label{eq_ham2}
H = -\frac{1}{N} L_x^2 + \lambda(N_{zz} - N_{yy}),
\end{equation}
with $\lambda = q/2|c_2|$.
Since the commutator $[H, L_x^2 + N_{yz}^2 + (N_{zz} - N_{yy})^2]$ vanishes and $\braket{L_x^2 + N_{yz}^2 + (N_{zz} - N_{yy})^2} \simeq (N/2)(N/2+1)$, we can take a semi-classical approximation by considering the Hamiltonian Eq. (\ref{eq_ham2}) on a sphere with radius $N/2$ \cite{leyvraz2005},
\begin{equation}
\label{ }
H = \frac{N}{2}(-\lambda\sin\theta\cos\varphi - \frac{1}{2}\cos^2\theta),
\end{equation}
with polar angle $\theta$ and azimuthal angle $\varphi$, which give $L_x = N\cos\theta/2$, $N_{yz}=N\sin\theta\sin\varphi/2$, and $N_{zz}-N_{yy}=-N\sin\theta\cos\varphi$. Setting $\mu = \cos\theta$, this Hamiltonian can be further simplified into,
\begin{equation}
\label{eq_hamK1}
K \equiv \frac{2H}{N} = -\lambda\sqrt{1 - \mu^2}\cos\varphi - \frac{1}{2}\mu^2,
\end{equation}
satisfying the Poisson bracket $\{ \mu, \varphi\} = 2/N$. For low-lying states,
we expand $K$ around its minimum which is found from
\begin{equation}
\label{ }
\sin\varphi_0 = 0,~{\rm and}~\mu_0 = \left\{\begin{array}{c}0,~~~~~~~~~~~~~~~~(\lambda>1) \\ \pm\sqrt{1-\lambda^2},~~~~(\lambda\leqslant1)\end{array}\right. ,
\end{equation}
and the corresponding minimum value of $H$ is
\begin{equation}
\label{ }
H_0 = \left\{\begin{array}{c}-\frac{\lambda N}{2}, ~~~~~~~~~~~ (\lambda>1) \\ -\frac{N}{4}(1 + \lambda^2),~~(\lambda\leqslant1)\end{array}\right. .
\end{equation}
Focusing on the broken-axisymmetry (BA) phase region ($\lambda \leqslant 1$), and expanding around the minimum of $\mu = \mu_0$ and $\varphi = 0$, and keeping the lowest order terms, the Hamiltonian (\ref{eq_hamK1}) becomes,
\begin{equation}
\label{eq_hamK2}
K = -\frac{1}{2} - \frac{\lambda^2}{2} + \frac{1 - \lambda^2}{2\lambda^2}\epsilon^2 - \frac{\lambda^2}{2}\varphi^2
\end{equation}
for low-lying levels, where $\epsilon = \mu - \mu_0$. If $\mu$ and $\varphi$ are taken as canonical conjugate variables for
`position' and `momentum', satisfying the commutator $[\mu, \varphi] = i\hbar$ according to Poisson bracket,
the harmonic spectra is obtained from Hamiltonian (\ref{eq_hamK2}) with frequency $\omega = \sqrt{1 - \lambda^2}$ and mass $m = 1/\lambda^2$. 
We note that the undepleted approximation breaks down as $q$ ramps down into the BA phase, especially when approaching to our target Dicke state $(\braket{N_0} = N/2)$. However, such a simple harmonic picture works well for the low-lying levels, as the comparison between the mean-field spectrum and the results from exact diagonalization shown in Fig. \ref{fig_sup6}. Therefore, this approximate tunable harmonic model remains capable of capturing the basic physics insight in the third stage of the DRL profile. During the third stage, the DRL profile simply shifts the Gaussian wave packet right after crossing the QCP at $\mu_0 \approx 0$ ($q/|c_2| \approx 2$ or $\lambda \approx 1$) to $\abs{\mu_0} \simeq 1$ ($q\simeq0$ or $\lambda \simeq0$).

\section{Deteriorating Dicke state quality with atom loss} \label{sec8}
\begin{figure}[!htp]
\begin{center}
\includegraphics[width=0.8\textwidth]{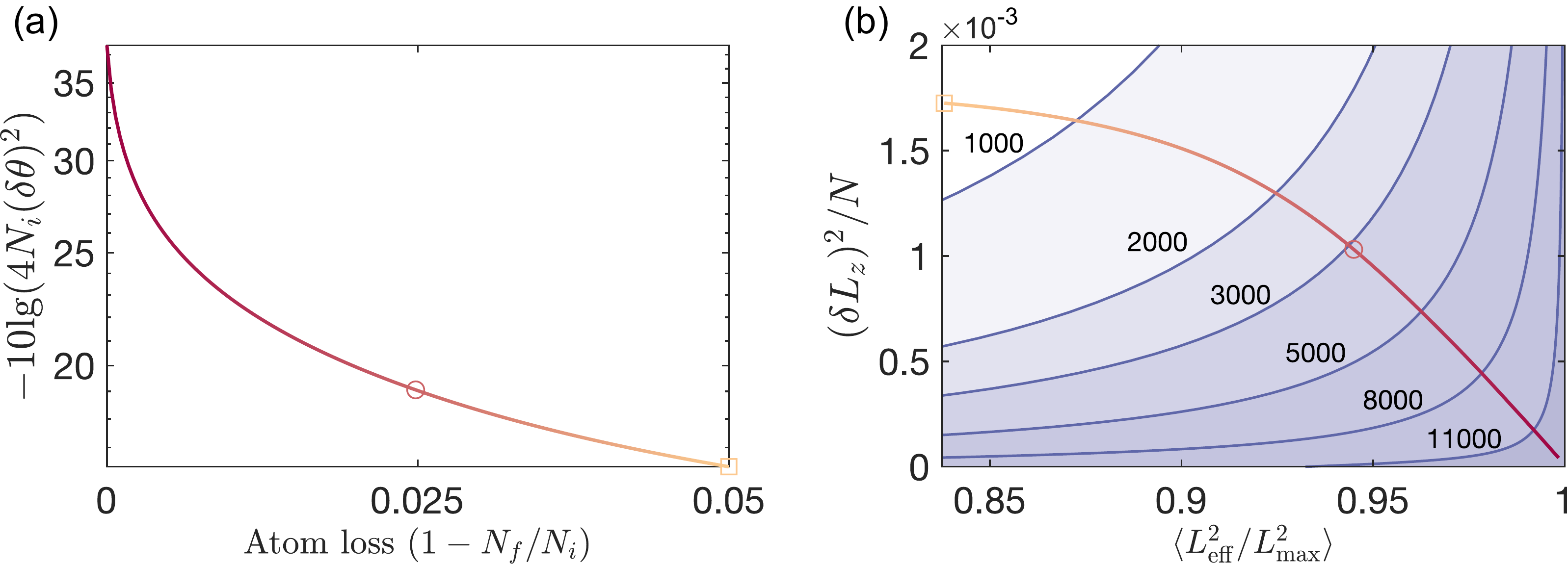}
\caption{Numerically calculated deteriorating Dicke state quality due to atom loss for $N = 11800$
using the $N=2000$ DRL policy with fidelity as reward (for without atom loss).
(a) The corresponding interferometric sensitivity $(\delta\theta)^2$, and (b) entanglement depth. The squares and circles mark the locations for 5\% and 2.5\% atom lost respectively.}
\label{fig_sup7}
\end{center}
\end{figure}

In our experiments, both detection noise $\sigma_{\rm DN}$ and atom loss degrade the observed final Dicke state quality. This section addresses the influence of atom loss by varying the loss rate $\gamma$ in the simulation with $N_f = N_i\exp(-\gamma \tau)$ and using the $N = 2000$ DRL policy $\pi^*({\cal A|S})$ with fidelity as reward
for $N = 11800$, since detection noise is well understood and can be directly calibrated.
The simulated interferometric sensitivity deterioraion is shown in Fig. \ref{fig_sup7}(a)
and analogously for entanglement depth of the final state \cite{sorensen2001,lucke2014,vitagliano2017} shown in Fig. \ref{fig_sup7}(b).
The empirical protocol adopted previously \cite{zou2018} demands 1.5 s to complete the sweep, which
results in $5\%$ loss of atoms (marked by a cross in Fig. \ref{fig_sup7}).
At $N = 11800$ and including atom loss, the sweep time for the DRL profile we find shortens to 766 ms with the corresponding atom loss reduced to $2.5\%$ (marked by a circle in Fig. \ref{fig_sup7}).
As shown in Fig. \ref{fig_sup7}, this reduction in atom loss leads to about 2.5 dB improvement to sensitivity and the number of entangled atoms at one standard deviation confidence level increased from less than 1000 to about 3000 atoms.

\section{sensitive dependence on parameters of DRL profile} \label{sec9}
As mentioned in section \ref{sec2}, due to the favorable generalization ability, the DRL agent trained at $N = 2000$ without loss can be directly applied to a system of $10^4$ atoms to achieve a final state fidelity ${\cal F}(\tau) > 0.8$. For our experiment carried out at $N = 11800$, such a DRL policy gives a target state fidelity ${\cal F} \simeq 0.85$ within $|c_2|\tau = 24$ (about 1500 ms for $c_2 = -2\pi \times 2.7$ Hz), which can facilitate an interferometric sensitivity of $-10\lg((\delta\theta)^2/(1/4N)) \simeq 37.7 $ dB, approaching Heisenberg limit. In contrast, the profile learnt from environment including loss discussed in the main text with $|c_2|\tau = 13$ only provides $-10\lg((\delta\theta)^2/(1/4N)) \simeq 22.1$ dB improvement in theory.

In a real experiment, it is always helpful to take into account the sensitive dependence on parameters
before the theoretical DRL sweeping profiles are implemented. According to the analysis in the main text, the third stage of our DRL profile corresponds formally to a STA problem of
translating a wave packet of a simple harmonic oscillation in the Fock basis.
According to previous studies, such a translational
displacement is susceptible to excitation of dipole (sloshing) mode \cite{couvert2008, ness2018}
using a STA.
For a Gaussian wave packet in a perfect harmonic trap, the residual amplitude of the dipole mode is given by \cite{couvert2008, ness2018},
\begin{equation}
\label{ }
{\cal A} = \abs{\int_0^{|c_2|\tau}\exp(-i\omega t)\dot{q}(t){\rm d}t},
\end{equation}
which depends on the sweeping time $|c_2|\tau$ and the profile $q(t)$. However, the coupling strength $|c_2|$ fluctuates from experiment to experiment due to uncertainty of the initial atom numbers (shot-noise in BEC
state preparation) and decays due to atom loss. In the end, for any chosen implementation of
sweep profile $q(t)$, variations of $|c_2|$ result in dipole mode excitations,
which finally lead to (averaged) deteriorating interferometric performance.

\begin{figure}[!htp]
\begin{center}
\includegraphics[width=\textwidth]{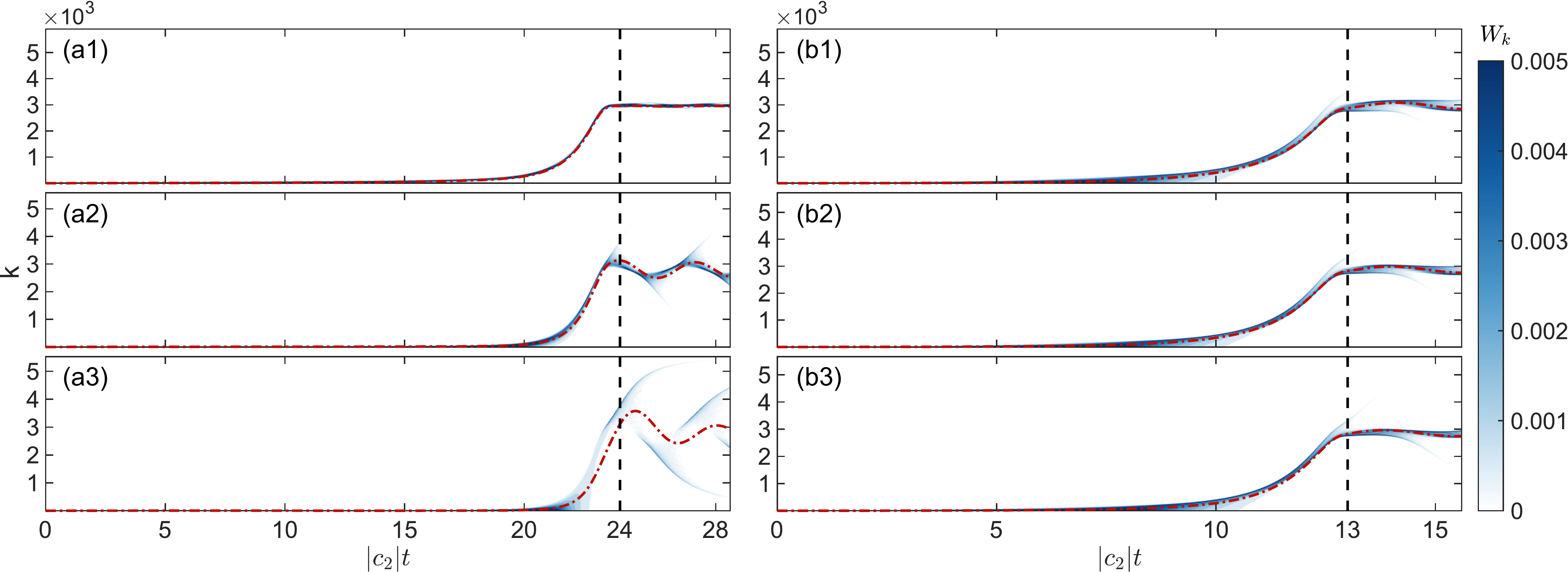}
\caption{Calculated evolution of excitation spectra $W_k = \abs{\braket{\psi(t)|k}}^2$ in the Fock basis $\ket{k,N-2k,k}$. (a1), (a2), and (a3) denote results from the high-fidelity profile from without loss for 11800 atoms, respectively for a constant $|c_2|$, a decaying $|c_2(t)|$, and a decaying $|c_2(t)|$ with initial atom number deviation
(${N} \to N - \sqrt{N}$). (b1), (b2) and (b3) represent the corresponding results from the learnt DRL profile including loss for $11800$ atoms. The red dashed-dotted lines denote the mean position in the Fock basis and the black dashed lines represent end of the sweep, after which $q$ is fixed to 0 ($t > \tau$).}
\label{fig_sup8}
\end{center}
\end{figure}

In the following, we compare numerically the high-fidelity policy profile from without loss with
the one from including loss, taking into account for both decaying $|c_2|$ and
shot-noise on initial atom numbers in Fig. \ref{fig_sup8}. Based on the residual sloshing motion
in the Fock basis after sweep, we find the profile from without loss is sensitive to variation of $|c_2|$ (see Figs. \ref{fig_sup8}(a2) and (a3)), while the one including loss is relatively insensitive (see Figs. \ref{fig_sup8}(b2) and (b3)). Thus, the profile trained under more realistic environment
including loss is experimentally feasible, and it is chosen for the reported experiments.

\section{experimental method} \label{sec10}
Although the sensitivity to atom loss has been partially eliminated after training the system including atom loss, the DRL profile is sensitive to the fluctuation $\Delta q_t$ and offset $\Delta q$ of the quadratic Zeeman shift. For the typical experimental system size ($N\simeq11800$), our simulation shows that the fluctuation of $\Delta q_t/|c_2|\in [-0.05,0.05]$ from the DRL profile would have negligible effects on the spin-mixing dynamics. Assuming a constant offset $\Delta q/|c_2| \in [-0.2,0.03]$ results in more significant effects and would lead to a 5\% reduction in the resulting phase sensitivity $(\delta \theta)^2$ under our current settings. 
The effect of positive offset ($\Delta q>0$) is much larger than its opposite situation, since a positive offset prolongs the stay in the polar phase instead of crossing the QCP into the BA phase.

To achieve the required stability, we have set up feedback-control systems for stabilizing both of the microwave power and magnetic fields which determine $q$. The microwave sensor and magnetic flux gates are both temperature stabilized, in order to inhibit drift of detection efficiency. The shot-to-shot fluctuation of our microwave power is around 0.1\% over a 12-hour period, which leads to $\Delta q_{\rm MW}/|c_2| \leqslant 0.02$. Our magnetic field control system has a bandwidth of $\sim$1 kHz, which can suppress peak-to-peak field fluctuation to $150~\mu$G, leading to $\Delta q_{\rm B}/|c_2|<0.008$ of negligible effect. In addition, we also have compensated the small magnetic field gradient around BEC with specially designed coils, since the gradient can cause phase separation of spin-up and spin-down components and therefore gives rise to to reduced overlap of different components as well as drift of spin exchange rate $|c_2|$.

In our experiment, a BEC with about $8.3 \times 10^4$ atoms is first produced in the $m_F = -1$ state inside an optical dipole trap formed by two crossed 1064-nm light beams in our system \cite{luo2017, zou2018}.
The quantization axis is defined by applying a fixed magnetic field of 815 mG along the direction of gravity, which gives $q_B/|c_2| \simeq 17$.
A RF $\pi/2$-pulse is then applied to transfer atoms from $m_F =-1$ to the $m_F=0$ component and a gradient magnetic field about 200 G/cm is ramped up to remove the remaining atoms in the $m_F = \pm 1$ components.
The remaining atom number in the $m_F=0$ component can be flexibly controlled via changing intensity of the RF pulse.
After that, the power of trapping light beams is lowered in 500 ms to further evaporate, while the gradient magnetic field is switched off.
Next, the trap is hold for another 500 ms followed by compression to the final trapping frequencies of $2\pi \times (237, 112,183)$ Hz along three orthogonal directions in 300 ms to produce a condensate of 11800 atoms.
During the last 300 ms, $q/|c_2|$ is ramped from $\sim 17$ to $3$ by linearly tuning up the dressing microwave.
This prepares a BEC sample for subsequent experiment to generate balanced Dicke state.

The application of DRL profile starts from abruptly jumping $q$ to the staring value $q(0)/|c_2| < 2$ and following the DRL sweep profile for 766 ms.
A 8 ms Stern-Gerlach separation is then applied after switching off the trap to obtain normalized atom numbers in each $m_F$ components by taking absorption images.
To measure the effective spin length, a RF-resonant $\pi/2$-pulse coupling $\ket{F =1,m_F =0}$ and $\ket{F =1,m_F =\pm1}$ is applied to rotate the spin-1 Dicke state. The same Stern-Gerlach process is then implemented after rotation to obtain normalized atom numbers.

The calibration of $q$ and $|c_2|$ gives $|c_2| = 2\pi \times 2.7$ Hz \cite{luo2017, zou2018},
at which experimental results match numerical simulations with the same value.


\begin{thebibliography}{60}
\bibitem [{sta()}]{state_space}%
  \BibitemOpen
  \href@noop {} {}\bibinfo {note} {$\braket{N_0}/N$ and $\arg\braket{
  {a}_{+1}^{\dagger} {a}_{-1}^{\dagger} {a}_{0}^2}$ are directly related to the
  two mean-field valuables for describing the semi-classical dynamics of the
  system, which is unfortunately insufficient to describe complete QPT. Hence
  we choose to augment with two observables describing the first-order quantum
  fluctuations. Both of these additional terms,
  $|\braket{a_{+1}^{\dagger}a_{-1}^{\dagger}a_0^2}|/N^2$ and $\braket{\delta
  N_0^2}/N^2$, come from the spin exchange interaction in the Hamiltonian
  (1).}\BibitemShut {Stop}%
\bibitem [{cpu()}]{cpu_info}%
  \BibitemOpen
  \href@noop {} {}\bibinfo {note} {The training process in this work is carried
  on two Intel Xeon Gold-6154 CPU (3.0 GHz) processors.}\BibitemShut {Stop}%
\bibitem [{err()}]{erratum}%
  \BibitemOpen
  \href@noop {} {}\bibinfo {note} {In the process of rechecking earlier
  experiments, we discover an error for the uncertainty of number squeezing
  reported in [9] 0.45 dB (1.48 dB after subtracting the detection noise),
  which overestimates the fluctuation by a factor of $\ln(10)$. The revised
  uncertainty of number squeezing is 0.20 dB (0.63 dB after subtracting the
  detection noise). For comparison, when the previous empirical sweep profile
  is implemented again under current experimental conditions ($N \simeq
  11800$), we measure $(\delta L_z)_{\theta=0} = 25.82 \pm 0.59$ from $1000$
  consecutive samples (vs. $25.51\pm 0.55$ for slightly different $N$ and the
  same number of experimental runs) with a number squeezing of $\xi^2_N =
  -20\log_{10}[(\delta L_z)_{\theta=0}/\sqrt{N}]\simeq 12.21 \pm 0.21~{\rm dB}$
  below the quantum shot noise of $\sqrt{N} \simeq 105.23 \pm 0.99$ for the
  polar state. The corresponding detection-noise-subtracted number squeezing is
  $17.34 \pm 0.65$ dB.}\BibitemShut {Stop}%
\end{thebibliography}
\end{document}